\documentclass[lineno]{jfm}

\usepackage{graphicx}
\usepackage{newtxtext}
\usepackage{newtxmath}
\usepackage{natbib}
\usepackage{mathtools}
\usepackage{hyperref}
\usepackage{tikz}
\usepackage{cleveref}
\hypersetup{
    colorlinks = true,
    urlcolor   = blue,
    citecolor  = blue,
}

\newcommand{\AxisRotator}[1][rotate=0]{%
    \tikz [x=0.25cm,y=0.7cm,line width=.1ex,-stealth,#1] \draw (0,0) arc (-150:150:1 and 1);%
}

\newcommand{\wT}{\mbox{$\smash{\mean{wT}}$}}

\newcommand{\volav}[1]{\left\langle #1 \right\rangle}

\newcommand{\horav}[1]{\langle #1 \rangle_h}
\newcommand{\timeav}[1]{\overline{#1}}
\newcommand{\mean}[1]{\timeav{\volav{#1}}}

\newcommand{\Tspace}{\mathcal{H}}

\newcommand{\lm}{\lambda} 
\newcommand{\bp}{\beta} 

\newcommand{\subeqref}[2]{\hyperref[#1]{(\ref*{#1}#2)}}

\DeclareMathOperator*{\esssup}{ess\,sup}

\definecolor{matlabblue}{RGB}{0,113,188}
\definecolor{matlabred}{RGB}{216,82,24}
\definecolor{mygrey}{rgb}{0.6,0.6,0.6}
\definecolor{matlabgreen}{rgb}{0,0.498,0} 
\definecolor{black}{rgb}{0,0,0}
\definecolor{mypurple}{rgb}{0.494,0.184,0.556}
\definecolor{colorbar1}{rgb}{1.000000,0.909091,0.000000}
\definecolor{colorbar2}{rgb}{1.000000,0.818182,0.000000}
\definecolor{colorbar3}{rgb}{1.000000,0.727273,0.000000}
\definecolor{colorbar4}{rgb}{1.000000,0.636364,0.000000}
\definecolor{colorbar5}{rgb}{1.000000,0.545455,0.000000}
\definecolor{colorbar6}{rgb}{1.000000,0.454545,0.000000}
\definecolor{colorbar7}{rgb}{1.000000,0.363636,0.000000}
\definecolor{colorbar8}{rgb}{1.000000,0.272727,0.000000}
\definecolor{colorbar9}{rgb}{1.000000,0.181818,0.000000}
\definecolor{colorbar10}{rgb}{1.000000,0.090909,0.000000}
\definecolor{colorbar11}{rgb}{1.000000,0.000000,0.000000}
\definecolor{colorbar12}{rgb}{0.909091,0.000000,0.000000}
\definecolor{colorbar13}{rgb}{0.818182,0.000000,0.000000}
\definecolor{colorbar14}{rgb}{0.727273,0.000000,0.000000}
\definecolor{colorbar15}{rgb}{0.636364,0.000000,0.000000}
\definecolor{colorbar16}{rgb}{0.545455,0.000000,0.000000}
\definecolor{colorbar17}{rgb}{0.454545,0.000000,0.000000}
\definecolor{colorbar18}{rgb}{0.363636,0.000000,0.000000}
\definecolor{colorbar19}{rgb}{0.272727,0.000000,0.000000}
\definecolor{colorbar20}{rgb}{0.181818,0.000000,0.000000}
\definecolor{colorbar21}{rgb}{0.090909,0.000000,0.000000}

\definecolor{matlabyellow}{rgb}{0.93,0.69,0.13} 

\newcommand\solidrule[1][10pt]{\rule[0.5ex]{#1}{1.5pt}}
\newcommand\dashedrule{\mbox{\solidrule[2pt]\hspace{2pt}\solidrule[2pt]\hspace{2pt}\solidrule[2pt]}}
\newcommand\dotdashedrule{\mbox{\solidrule[3pt]\hspace{1.5pt}\solidrule[1pt]\hspace{1.5pt}\solidrule[3pt]}}

\newcommand{\mysquare}[1]{%
	\protect\begin{tikzpicture}%
	\protect\fill [color=#1] (0,0) -- (0.75ex,0) -- (0.75ex,0.75ex) -- (0,0.75ex) -- (0,0);
	\protect\end{tikzpicture}%
}

\newcommand{\mycirc}[1]{%
	\protect\begin{tikzpicture}%
	\protect\draw[thick,color=#1] (0.5ex,0.5ex) circle (0.5ex);
	\protect\end{tikzpicture}%
}

\renewcommand{\vec}[1]{\mathbf{#1}}

\renewcommand{\theta}{\vartheta}
\renewcommand{\phi}{\varphi}

\newcommand{\bR}{\mathbb{R}}        

\usepackage[normalem]{ulem}

\newtheorem{lemma}{Lemma}

\newcommand{\RomanNumeralCaps}[1]

\title{
\nolinenumbers 
Internally heated convection with rotation: bounds on heat transport}

\author{ 
\nolinenumbers 
Ali Arslan
  \corresp{\email{ali.arslan@erdw.ethz.ch}}}

\affiliation{ 
\nolinenumbers
Institute of Geophysics, ETH Z\"{u}rich, Z\"{u}rich, CH-8092, Switzerland
}

\begin{document}
\maketitle

\begin{abstract}
\nolinenumbers
\noindent
This work investigates heat transport in rotating internally heated convection, for a horizontally periodic fluid between parallel plates under no-slip and isothermal boundary conditions. The main results are the proof of lower bounds on the mean temperature, $\mean{T}$, and the heat flux out of the bottom boundary, $\mathcal{F}_B$, at infinite Prandtl number, where the Prandtl number is the nondimensional ratio of viscous to thermal diffusion. The lower bounds are functions of the Rayleigh number quantifying the ratio of internal heating to diffusion and the Ekman number, $E$, which quantifies the ratio of viscous diffusion to rotation. We utilise two different estimates on the vertical velocity, $w$, one pointwise in the domain \citep[][\emph{J. Math. Phys.}, vol.45(7), pp.2718-2743]{Yan2004} and the other an integral estimate over the domain \citep[][\emph{Phys. D: Non. Phen.}, vol. 125, pp. 275-284]{constantin1999r}, resulting in bounds valid for different regions of buoyancy-to-rotation dominated convection.
Furthermore, we demonstrate that similar to rotating Rayleigh-B\'enard convection, for small $E$, the critical Rayleigh number for the onset of convection asymptotically scales as $E^{-4/3}$. 

\end{abstract}


\nolinenumbers

\section{Introduction}
\label{sec:intro}

Heat transport by turbulent convection remains a pertinent area of research in both astrophysical and geophysical fluid dynamics. 
While boundary-forced thermal convection has been extensively studied, convection driven by internal heating has been relatively overlooked \citep{doering2020turning}. Nevertheless, internally heated convection (IHC) plays a significant role within planetary bodies, such as in the Earth's mantle and core, where the radioactive decay of isotopes and secular cooling drive fluid motion \citep{schubert2001mantle,schubert2015treatise}. Similarly, for stars, convective zones are driven by radiation from nuclear fusion \citep{schumacher2020} and supernovae are modelled as fluids heated internally by neutrinos \citep{herant1994,radice2016}. Moreover, stars and planets are rotating bodies where the Coriolis force significantly affects the flow dynamics \citep{greenspan1968}.

Studying rotating turbulent convection is challenging because experiments and numerical simulations cannot reach parameter values of interest \citep{glatzmaier2013}. For example, in planetary mantles, the Prandtl number, $Pr$, the non-dimensional number quantifying the ratio of the viscous and thermal diffusivity, reaches values of $10^{23}$, while the Rayleigh number, $R$, quantifying the ratio of thermal forcing to diffusion is at least $10^6$ \citep{mulyukova2020}. In planetary cores, the Rayleigh number could be as high as $10^{26}$ \citep{schubert2015treatise}. Furthermore, the Ekman number, $E$, representing the viscous to rotational forces, is estimated to be $10^{-15}$ in the Earth's core \citep{jones2015}.

\begin{figure}
    \centering
    \begin{tikzpicture}[every node/.style={scale=0.95}]

    \draw [-stealth,colorbar15, thick] (-0.8,0) -- (-0.8,-0.5);
    \draw [-stealth,colorbar15, thick]  (0.2,0) -- (0.2,-0.5);
    \draw [-stealth,colorbar15, thick] (1.2,0) -- (1.2,-0.5);
    \draw [-stealth,colorbar15, thick]  (2.2,0) -- (2.2,-0.5);
    \draw [-stealth,colorbar15, thick]  (-0.8,2.3) -- (-0.8,3);
    \draw [-stealth,colorbar15, thick]  (0.2,2.3) -- (0.2,3);
    \draw [-stealth,colorbar15, thick]  (1.2,2.3) -- (1.2,3);
    \draw [-stealth,colorbar15, thick]  (2.2,2.3) -- (2.2,3);
    \draw[black,very thick, fill=mygrey, fill opacity = 0.2] (-2,2) -- (-1,2.6) -- (3.3,2.6) -- (2.3,2) -- cycle;
    \draw[black,very thick, fill=mygrey, fill opacity = 0.2] (-2,0) -- (-1,0.6) -- (3.3,0.6) -- (2.3,0) -- cycle;
    \draw[->] (-2.6,1.5) -- (-2.6,1) node [anchor=north] {$g$};
    \draw[->] (-1.6,0.1) -- (-1,0.1) node [anchor=west] {$x$};
    \draw[->] (-1.6,0.1) -- (-1.1,0.4) node [anchor=west] {$y$};
    \draw[->] (-1.6,0.1) -- (-1.6,1.6) node [anchor=south] {$z$};
    \node at (-1.6,1) {\AxisRotator[rotate=-90]};
    \node at (-2.1,0.4) {$z=0$};
    \node at (-2.1,2.4) {$z=1$};
    \node at (3.6,0.3) {$ T = 0 $};
    \node at (3.6,2.3) {$  T = 0 $};
    \node at (0.7,-0.6) {${\color{colorbar15}\mathcal{F}_B }  $};
    \node at (1.8,1.3) {${\color{colorbar15}\mean{T}}  $};
    \node at (0.7,3) {${\color{colorbar15}\mathcal{F}_T }  $}; 
    \node at (0.3,1.3) {$ H = 1$};
    
    \end{tikzpicture}
    \caption{ A non-dimensional schematic diagram for rotating uniform internally heated convection. The upper and lower plates are at the same temperature, and the domain is periodic in the $x$ and $y$ directions and rotates about the $z$ axis. $\mathcal{F}_B$ and $\mathcal{F}_T$ are the mean heat fluxes out the bottom and top plates, $\mean{T}$ the mean temperature, and $g$ is the acceleration due to gravity. }
    \label{fig:config}
\end{figure}
An alternative route for inquiry is a mathematically rigorous study of the equations describing rotating convection. Of interest is the regime where the solutions of the governing equations are turbulent, and a key question is on the long-time behaviour of the mean quantities of the flow as a function of the control parameters ($Pr, R, E$). In this study, we employ the \textit{background field method} \citep{doering1992energy,Doering1994,Constantin1995a,Doering1996} to study the mean heat transport in IHC subject to rotation between parallel plates with isothermal and no-slip boundary conditions (\cref{fig:config}). Unlike turbulent convection driven by boundary heating, i.e. Rayleigh-B\'enard convection (RBC), there are no known rigorous results for turbulent IHC subject to rotation.

The influence of rotation alters turbulent convection and introduces new flow regimes and physics (see \cite{ecke2023} for a recent review). The flow features of rotating convection in a plane layer driven by boundary heating are well documented \citep{chandrasekhar2013hydrodynamic,veronis1959,rossby1969,boubnov2012,julien1996,knobloch1998,vorobieff2002,stevens2013}, and some insight exists for non-uniform IHC \citep{barker2014,currie2020,hadjerci2024}. However, no study has explored the flow in rotating uniform IHC. The preceding studies show that rotation inhibits the onset of convective motion and stabilises the fluid, creating a bias in motion parallel to the axis of rotation. Further, an Ekman boundary layer exists, enhancing the mean vertical heat transport by Ekman pumping \citep{greenspan1968}. With sufficient thermal forcing, the $E-R$  parameter space contains two extreme flow states: if $R$ is sufficiently larger than $E^{-1}$, then buoyancy dominates and rotation plays little effect on the dynamics, whereas if $E^{-1}$ is large relative to $R$, and the vertical velocity is nonzero, geostrophic turbulence occurs \citep{julien1996,sprague2006}. A wide range of flow features occurs in rotating convection including, cellular flows, Taylor columns, large-scale vortices and plume-dominated convection \citep{grooms2010, julien2012, stellmach2014, guzman2020,aurnou2020,kunnen2021,Song2024}.

In addition to experimental and numerical studies on rotating RBC, there exist proofs of bounds with the background field method, on the enhancement of heat transport due to convection, quantified with the Nusselt number, $Nu$ \citep{constantin1999r,Doering2001,constantin2001,Yan2004,grooms2014,Pachev2020}. First introduced in the 1990s, the background field method provides a tool for proving bounds on the long-time averages of turbulence \citep{Fantuzzi2022}. In its original formulation, the idea involves decomposing the flow variables into a fluctuating and background component satisfying the boundary conditions to construct a variational problem for bounding the turbulent dissipation. A bound is proven by solving the variational problem by choosing an appropriate background field and using elementary integral estimates.
The method has been successfully used for many fluid flows, none more so than turbulent convection \citep{Nobili2022}. Recent insight has shown that the background field method fits within the framework of the \textit{auxiliary functional method} \citep{Chernyshenko2014a,Chernyshenko2022}, which can yield sharp bounds for well-posed ODEs and PDEs under technical conditions \citep{Tobasco2018,Rosa2020}.

A fundamental feature of the background field method is to work with energy balances from the governing equations. However, energy identities fail to capture the effects of rotation, apart from in the case of a fluid driven by rotating boundaries, like in Taylor-Couette flow \citep{constantin1994,ding2019tc,Kumar_2022tc}. For convection subject to the Coriolis force, standard applications of the background field method do not give a bound on $Nu$ that depends on $E$. One path for progress is in the limit of infinite $Pr$, where the momentum equation simplifies to a forced Stokes flow, leading to a diagnostic equation between the velocity and temperature, facilitating better estimates. Notably, without rotation ($E=\infty$), using the background field method, it was proven, up to constants and logarithms, that $Nu\leq Ra^{1/3}$ \citep{Doering2006}, where $Ra$ is the Rayleigh number based upon the temperature difference between the boundaries, improving on the bound of $Nu\leq Ra^{1/2}$ valid at arbitrary $Pr$ \citep{Doering1996}. Under rotation ($E<\infty$) at $Pr=\infty$, established results for RBC are illustrated in \cref{fig:upper_bounds_RRBC}.

\begin{figure}
    \centering
    \begin{tikzpicture}[every node/.style={scale=0.95}]
    \draw[->,black,very thick] (-7,0) -- (5,0) ;
     \draw[dotted] (-6.4,4.5) -- (5,4.5);
    \draw[->,black, very thick] (-6.6,-0.4) -- (-6.6,5);
    \node at (5.5,0) {$E$};
    \node at (-7,4.8) {$Nu$};
    \draw[black,dotted] (-6.4,2) -- (5,2);
    \draw[matlabblue,thick] (2,2) -- (5,2);
    \node at (3.6,1.5) {\scriptsize$ {\color{matlabblue} (Ra \ln \ln Ra)^{1/3} } $};
    \node at (3.6,1.1) {\scriptsize${\color{matlabred}Ra^{5/12}}$};
    \draw[matlabblue,thick] (-4,4.5) -- (-2,4.5);
    \draw[matlabgreen,thick,dashed] plot[smooth,tension=1] coordinates{ (-6.6,0)  (-4,2) (-2.5,4.5)};
    \node at (-2.9,4.8) {\scriptsize${\color{matlabblue} Ra^{2/5} } $};
    \node at (-0.5,3.7) {\scriptsize$\color{matlabblue}{Ra^{4/11}}$};
    \draw[matlabblue] (-0.9,3.6) -- (-1.7,3) ;
    \draw[matlabblue,thick] (-2,3) -- (-1.2,3);
    \draw[dotted] (-1,3) -- (5,3);
    \draw[matlabblue,thick] (-6.6,0) -- (-4,4.5);
    \draw[matlabblue,thick] plot[smooth,tension=1] coordinates{(-3.5,4.5) (-2.9,3.7) (-2,3)};
    \draw[matlabblue,thick] plot[smooth,tension=1] coordinates{(-2,4.5) (-0.8,2.7) (2,2)};
    \node at (-5.7,2.8) {\scriptsize${\color{matlabblue} E\, Ra^2 } $};
    \node at (-5.7,3.2) {\scriptsize${\color{matlabred}E^2 Ra^2}$};
    \node at (1.2,2.7) {\scriptsize${\color{matlabblue}  E^{-4/5} Ra^{1/5} } $};
    \node at (-4,-0.2) {\tiny ${\color{matlabgreen}  E\leq Ra^{-2/3}  } $};
    \draw[black,dotted] (-4,-0.6) -- (-4,4.5);
    \draw[black,dotted] (-2,-0.6) -- (-2,4.5);
    \draw[black,dotted] (2,-0.3) -- (2,2);
    \draw[black,dotted] (-2.5,4.5) -- (-2.5,-0.5);
    \draw[<->,matlabgreen] (-6.4,-0.4) -- (-2.6,-0.4) ;
    \draw[<->] (-6.4,-0.5) -- (-4.2,-0.5) ;
    \draw[<->] (-3.8,-0.5) -- (-2.2,-0.5) ;
    \draw[<->] (-1.8,-0.2) -- (1.8,-0.2) ;
    \draw[<->] (2.2,-0.2) -- (4.8,-0.2) ;
    \draw[matlabblue] (-3,1.9) -- (-2.7,3.5) ;
    \draw[matlabblue] (-5.5,2.6) -- (-5.2,2.4) ;
    \draw[matlabblue] (-3,4.67) -- (-3,4.5) ;
    \draw[matlabblue] (1.1,2.5) -- (1,2.08) ;
    \draw[matlabblue] (3.5,1.7) -- (3.5,2) ;
    \draw[matlabgreen] (-4.1,2.3) -- (-3.9,2.1) ;
    \node at (-1.2,3) { \mysquare{matlabblue} };
    \node at (-3.5,4.5) { \mycirc{matlabblue} };
    \node at (-5,0.5) {$\textrm{A}$};
    \node at (-3,0.5) {$\textrm{B}$};
    \node at (0,0.5) {$\textrm{C}$};
    \node at (3.6,0.5) {$\textrm{D}$};
    \node at (-3,1.7) {\scriptsize $\color{matlabblue}{Ra^{\frac{4}{11}}(E^{-1} +1)^{\frac{4}{11}} }$};
    \node at (-4.1,2.8) {\scriptsize $\color{matlabred}{ E^{4} Ra^{3} }$};
    \node at (-4.1,2.5) {\scriptsize $\color{matlabgreen}{ E^{2} Ra^{2} }$};
    \node at (-5.4,-0.8) {\tiny $E \leq\frac{1}{Ra^{2}} $};
    \node at (-3,-0.8) {\tiny $ \frac{1}{Ra^{2}} \leq E \leq \frac{1}{Ra^{\frac14} }  $};
    \node at (0,-0.6) {\tiny $ \frac{1}{Ra^{\frac14}} \leq E \leq \frac{1}{Ra^{\frac16}(\ln \ln Ra)^{\frac{5}{12} } } $};
    \node at (3.5,-0.6) {\tiny $ \frac{1}{Ra^{\frac16}(\ln \ln Ra)^{\frac{5}{12} } } \leq E $};
    \node at (-6.6,-0.7) {$0$};
    \node at (-7.2,0) {$1$};
    \end{tikzpicture}
    \caption[upper bounds]{Illustration of the upper bounds with blue lines ({\color{matlabblue}\solidrule}) on the Nusselt number, $Nu$ , in terms of the Rayleigh, $Ra$, and Ekman numbers, $E$, for infinite Prandtl number rotating Rayleigh-B\'enard convection, between no-slip boundaries. All bounds hold up to constants that determine the exact sizes of the regions and where applicable red text shows the stress-free result in the same region. The bounds in A are due to \cite{constantin1999r}. The green dashed line ({\color{matlabgreen}\dashedrule}) crossing A and B show the upper bound obtained from the non-hydrostatic quasigeostrophic approximation \citep{Pachev2020,grooms2014}. In B, the crossover from the bound $Ra^{2/5}$ \citep{Doering2001} to $Ra^{4/11}(E^{-1}+1)^{4/11}$ \citep{Yan2004} is shown by a blue circle (\mycirc{matlabblue}). Similarly in region C the crossover from $Ra^{4/11}$ \citep{Yan2004} to $E^{-4/5} Ra^{1/5}$ \citep{constantin2001} is shown by a blue square (\mysquare{matlabblue}). Region D shows the bounds for buoyancy-driven convection \citep{Otto2011,Whitehead2011prl}. The transition from C to D is continuous up to logarithmic corrections \citep{constantin1999r}.}
    \label{fig:upper_bounds_RRBC}
\end{figure}

High $Pr$ restricts the parameter space when modelling fluid flows. However, proving bounds in the limit of $Pr=\infty$ can be viewed as a first step towards establishing bounds valid for all $Pr$. Recent studies suggest that for any bound proven at infinite $Pr$ in rotating RBC, a semi-analytic bound for finite $Pr$ can be obtained under specific conditions \citep{tilgner2022}. The results in \cite{tilgner2022} indicate that the bounds for finite $Pr$ are, to highest order, equivalent to the infinite $Pr$ results of \cref{fig:upper_bounds_RRBC}, with $Ra, Pr$ and $E$ corrections. The result is unsurprising since bounds at infinite $Pr$ generally improve those obtained for finite $Pr$. At the level of the dynamical system, this can be understood as a consequence of the relative ease with which information is extracted from the turbulent attractor of infinite $Pr$ system by bounding methods \citep{wang2007asymptotic}.

When rotation dominates over buoyancy, heuristic arguments for RBC suggest that $Nu \sim E^{3/2} Ra^2$, at arbitrary $Pr$ \citep{king2013,plumley2019,aurnou2020}. Bounds that scale similarly to the physical arguments in the rapidly rotating regime can be proven when working with an asymptotic approximation of the governing equations known as the non-hydrostatic quasigeostrophic (nhQG) equations \citep{julien1996,julien2016}. Scaling the horizontal length scales by $E^{1/3}$ and adjusting the time variable yields the nhQG equations that model the limit of rapidly rotating convection in a plane layer. Applying the background field method to the nhQG equations gives the green bounds in \cref{fig:upper_bounds_RRBC} of, up to constants, $Nu\leq E^2 Ra^2$, for no-slip conditions \citep{Pachev2020} and $Nu \leq E^4 Ra^3$ for stress-free boundaries \citep{grooms2014}.

IHC remains less studied in part due to significant differences in the physics between RBC. Notably, in uniform IHC between isothermal boundaries, the mean conductive heat flux is zero, rendering the standard definition of the Nusselt number inapplicable \citep{Goluskin2016book}. In previous works with zero rotation \citep{Goluskin2016book}, an alternative measure of the turbulent convection is the non-dimensionalised mean temperature, $\mean{T}$, where angled brackets $\volav{\cdot}$ denote a volume and overbars a long-time average. As the flow becomes increasingly turbulent, the temperature within the domain becomes homogenised, quantified in a lower value of $\mean{T}$, and a higher proxy Nusselt number defined as $Nu_p  = 1/\mean{T}$. An additional measure of turbulence is $\wT$, quantifying the portion of heat leaving through each boundary, $\mathcal{F}_T$ and $\mathcal{F}_B$, due to convection \citep{goluskin2012convection}. For a stationary fluid, the heat supplied leaves the domain symmetrically out of both boundaries to ensure the statistical stationarity of the solutions. As the thermal forcing increases, convection carries heat upwards, causing a higher portion of the heat to leave through the top relative to the bottom boundary \citep{goluskin2016penetrative}.

In line with previous works on uniform internally heated convection \citep{Goluskin2016book,Arslan2021a,Arslan2021,kumar2021ihc,Arslan2023,Arslan2024}, the non-dimensional heat flux out of the top and bottom boundaries is given by 
\begin{subequations}
\label{eq:Flux_def}
    \begin{align}
    \mathcal{F}_T &:= -\mean{\partial_zT}_h\vert_{z=1} = \frac12 + \wT \quad \text{and}\\
    \mathcal{F}_B &:= \phantom{-}\mean{\partial_z T}_h\vert_{z=0} = \frac12 - \wT.
    \end{align}
\end{subequations}
The non-dimensionalisation sets the limits of $\wT$ as $0$ and $\tfrac12$, with each limit corresponding to no convection and infinitely effective convection, respectively. We seek bounds of the form, $\mathcal{F}_B \geq f_1(R,E)$ and $\mean{T}\geq f_2(R,E)$ in different regions of $E-R$ space, where $f_1$ and $f_2$ are functions of only $R$ and $E$. In previous applications of the background field method to IHC, bounds for $\mean{T}$ are proven with minor adaptation from the background field method as applied to RBC \citep{Lu2004,Whitehead2011,Whitehead2012}. However, in the case of obtaining bounds on $\wT$ and consequently $\mathcal{F}_B$, it has been established that the variational problem requires a minimum principle on $T$, which states that temperature in the domain is greater than or equal to zero \citep{Arslan2021}. The minimum principle is necessary to obtain lower bounds on $\mathcal{F}_B$ that remain positive as $R$ increases. In the case of no rotation at $Pr=\infty$, the best-known lower bound on the mean temperature, up to constants, are $\mean{T}\geq (R\ln{R})^{-1/4} $ \citep{Whitehead2011} and $\mean{T} \geq R^{-5/17}$ \citep{Whitehead2012} for no-slip and stress-free boundaries. Conversely, the best-known lower bounds on the heat flux out of the domain are $\mathcal{F}_B \geq R^{-2/3} + R^{-1/2}|\ln{(1-R^{-1/3})}|$ and $\mathcal{F}_B \geq R^{-40/29} + R^{-35/29} |\ln{(1-R^{-10/29})}|$ \citep{Arslan2024} for the two different kinematic boundary conditions.

In this paper, the bounds we prove for uniform internally heated convection subject to rotation at an infinite Prandtl number are summarised in \cref{tab:bound_summary}.
\begin{table}
    \centering
    \caption{Summary of the lower bounds on $\mathcal{F}_B$ and $\mean{T}$ proven in this work. The constants in the bounds are collated in \cref{tab:constants} for brevity, while $E_0 = 5.4927$, $E_1=8$ and $E_m = 41.4487$. Region I corresponds to buoyancy-dominated convection, II to solid body rotation, III to buoyancy or rotation-dominated convection and IV to rotation-dominated convection provided $R>R_L(E)$ where $R_L$ is the value of $R$ above which the system is linearly unstable.}
    \begin{tabular}{p{0.85cm}| p{1.8cm} | p{5.75cm} | p{2.25cm} | p{1.8cm}}
    \hline
         Region  & Condition & Bound on $\mathcal{F}_B$ & Condition & Bound on $\mean{T}$ \\ \hline
         I & $E\gtrsim R^{-2}$ & $ d_4 R^{-\frac23} + d_5 R^{-\frac12}\left|\ln(1- d_6 R^{-\frac13} )\right| $ & $E\gtrsim R^{-\frac23}$  &
        $d_{11} R^{-\frac27}  $\\ & $E \in [E_m,E_0]$& &$ E \in [E_m,E_1]$ & \vspace{2pt} \\ 
        Ia & & & $E \gtrsim R^{-\frac23}$ & $d_{13} R^{-1} E^{-1} $  \\ & & &$E \geq E_m$ &\vspace{10pt}\\
         II& $E\lesssim R^{-2}$ & $ \frac12 $  & $E\lesssim R^{-\frac23}$ &  $ \frac{1}{12} $ \\ & $E \in [E_m,E_0]$ &&  $E \in [E_m,E_1]$ & \vspace{10pt}\\
         III& $E\gtrsim R^{-2}$ & $ d_1\,R^{-\frac23}E^{\frac23} +  d_2\,R^{-\frac12}E^{\frac12} \left|\ln(1- d_3\, R^{-\frac13}E^{\frac13} )\right| $ & $E\gtrsim R^{-\frac59}$  & $ d_{10} R^{-\frac27}E^{\frac27} $  \\ & $E\leq E_0$ && $E\leq E_1$ & \vspace{2pt}\\ IIIa&  &  & $R^{-\frac59} \gtrsim E\gtrsim R^{-\frac23}$  & $ d_{13} R^{-1}E^{-1} $  \\ &  && $E\leq E_1$ & \vspace{10pt}\\
         IV&  $E\lesssim R^{-2}$& $ d_7 R^{-1} + d_8 R^{-\frac45}\left|\ln{(1- d_{9} R^{-\frac25})}\right| $ & $E\lesssim R^{-\frac23}$  & $ d_{12} R^{-\frac13}  $ \\ & $E\leq E_0$  && $E\leq E_1$ & 
    \end{tabular}
    \label{tab:bound_summary}
\end{table}


For notation, $\lVert f \rVert_p^p = \int^{1}_0 f^p \textrm{d}z$, for $p<\infty$ and $\lVert f \rVert_\infty = \esssup_{z\in [0,1]} f$ for $p=\infty$, represents the standard $L^p$ norms of $f:[0,1] \rightarrow \mathbb{R} $. The use of $\lesssim$ or $\gtrsim$ indicates equality up to an independent constant. The paper is structured as follows. In \cref{sec:setup}, we describe the problem setup before discussing the onset of convection in \cref{sec:stability}. Then \cref{sec:heuristics} proposes heuristic scaling arguments for rotating IHC before we prove bounds on $\mathcal{F}_B$ in \cref{sec:Bounds_wT} and on $\mean{T}$ in \cref{sec:Bounds_T}. Finally, \cref{sec:conclusion} offers a brief discussion and concluding remarks.

\section{Setup}
\label{sec:setup}

We consider a layer of fluid in a rotating frame of reference between two horizontal plates separated by a distance $d$ and periodic in the horizontal ($x$ and $y$) directions with periods $L_x d$ and $L_y d$.  The fluid has kinematic viscosity $\nu$, thermal diffusivity $\kappa$, density $\rho$, specific heat capacity $c_p$ and thermal expansion coefficient $\alpha$. Gravity acts in the negative vertical direction with strength $g$, the fluid rotates at rate $\Omega$ and is uniformly heated internally at a volumetric rate $H$. 

To non-dimensionalise the problem, we use $d$ as the characteristic length scale, $d^2/\kappa$ as the time scale and $d^2 H/\kappa \rho c_p$ as the temperature scale \citep{roberts1967convection}. The velocity of the fluid $\boldsymbol{u}(\boldsymbol{x},t) =u(\boldsymbol{x},t)\vec{e}_1 +  v(\boldsymbol{x},t)\vec{e}_2 +  w(\boldsymbol{x},t)\vec{e}_3$ and temperature $T(\boldsymbol{x},t)$ in the non-dimensional domain $V = [0,L_x]\times[0,L_y]\times[0,1]$ are governed by the infinite Prandtl number Boussinesq equations,
\begin{subequations}
\label{eq:gov_eqs}
\begin{align}   
    \label{eq:continuit}
    \bnabla \cdot \boldsymbol{u} &= 0,\\
    \label{eq:mom_eq}
    \bnabla p + E^{-1} \vec{e}_3 \times \boldsymbol{u} &=  \bnabla^2 \boldsymbol{u} +  R\, T \vec{e}_3 ,   \\
    \label{eq:nondim_energy}
    \partial_t T + \boldsymbol{u}\cdot \bnabla T &= \bnabla^2 T + 1.
\end{align}
\end{subequations}
The non-dimensional numbers are the Ekman and Rayleigh numbers, defined as 
\begin{equation}
    E = \frac{\nu}{2\Omega d^2},
    \qquad\text{and}\qquad 
    R = \frac{g \alpha  H d^5 }{\rho c_p \nu \kappa^2 }.
\end{equation}
The boundary conditions are of no-slip and isothermal temperature, respectively:
\begin{subequations}
\label{bc:non_uni}
\begin{gather}
    \label{bc:u_non_u}
    \boldsymbol{u}|_{z=\{0,1\}} = 0, \\
    \label{bc:nu}
    T|_{z=\{0,1\}} = 0\, .
\end{gather}
\end{subequations} 
Figure \ref{fig:config} provides a schematic for the system under consideration. The vertical component of the curl and double curl of \eqref{eq:mom_eq} gives a diagnostic equation involving the vertical velocity $w$, the vertical vorticity $\zeta$ and temperature $T$:
\begin{subequations}
\label{eq:mom_eqs_diag} 
\begin{gather}
\label{e:w_and_T_eq}
\bnabla^4 w = E^{-1} \partial_z \zeta -R \, \Delta_h T, \\
\label{e:z_w_eq}
\bnabla^2 \zeta = - E^{-1} \partial_z w,
\end{gather}
\end{subequations}
where $\Delta_h=\partial^2_x + \partial^2_y$ is the horizontal Laplacian.
 
The final ingredients are results from \cite{Yan2004} and \cite{constantin1999r} and a minimum principle on $T$. We state the results as separate lemmas.
\begin{lemma}[minimum principle]
    \label{lem:min}
    Suppose $T(\boldsymbol{x},t)$ solves  \eqref{eq:nondim_energy} subject to \eqref{bc:nu} where $\boldsymbol{u}$ satisfies $\bnabla\cdot \boldsymbol{u}=0$ and \eqref{bc:u_non_u}. Let the negative parts of $T(\boldsymbol{x},t)$ be
    \begin{equation}
        T_-(\boldsymbol{x},t) := \max\{-T(\boldsymbol{x},t),0\}. 
    \end{equation}
    Then
    \begin{equation}
        \volav{ |T_-(\boldsymbol{x},t)|^2} \leq \volav{ |T_{-}(\boldsymbol{x},0)|^2} \ \exp{(-\mu t)},
    \end{equation}
    for some $\mu > 0$. In particular, if $T(\boldsymbol{x},0) > 0$, then $T_{-}(\boldsymbol{x},0)=0$ and $T(\boldsymbol{x},t)\geq 0 ~ \forall t$. 
\end{lemma}
See the appendix A of \cite{Arslan2021} for a proof. 
\begin{lemma}[\cite{Yan2004}]
    \label{lem:yan}
    Let $w_{\boldsymbol{k}},T_{\boldsymbol{k}}:(0,1) \rightarrow \mathbb{R}$, be the Fourier transforms of the vertical velocity $w$ and temperature $T$ with wavenumber $\boldsymbol{k} = (k_x, k_y)$ satisfying \eqref{eq:mom_eqs_diag} and $\nabla \cdot \boldsymbol{u}=0$, subject to the velocity boundary conditions \eqref{bc:u_non_u}. Then, 
        \begin{itemize}
            \item For $|\boldsymbol{k}|\leq 1$
        \begin{equation}
        \label{eq:w_est_low_k}
        \lVert w''_{\boldsymbol{k}} \rVert_\infty \leq c_1 R (1+ \tfrac14 E^{-2})^{1/4} \lVert T_{\boldsymbol{k}} \rVert_2 ,
        \end{equation}
        where $c_1 = 6^{1/4}$.
        \item For $|\boldsymbol{k}| \geq 1$
        \begin{equation}
        \label{eq:w_est_high_k}
        \lVert w''_{\boldsymbol{k}} \rVert_\infty \leq c_2 R \sqrt{\boldsymbol{k}} \lVert T_{\boldsymbol{k}} \rVert_2 + c_2 R E^{-1} \lVert T_{\boldsymbol{k}} \rVert_2 ,
        \end{equation}
        where $c_2 = 1 + \frac{e^2 + 1}{e^2 - 1}\frac{4\cosh{1} + 2\sinh{1}}{-1 +\sinh{1}  } \sim 64.8734$.
        \end{itemize}
\end{lemma}
\begin{lemma}[\cite{constantin1999r}]
    \label{lem:const}
    Let $w,T: V \rightarrow \mathbb{R}$ be horizontally periodic functions such that they solve \eqref{eq:mom_eqs_diag} subject to $\nabla \cdot \boldsymbol{u}=0$ and the boundary conditions \eqref{bc:u_non_u}, then
    \begin{equation}
        \volav{  |\bnabla^2 w|^2  }  + 2 \volav{  |\bnabla \zeta|^2 } \leq R^2  \volav{|T|^2} .
    \end{equation}
\end{lemma}

\section{Onset of convection}
\label{sec:stability}

Before proving bounds on the emergent properties of the turbulence ($\mathcal{F}_B$ and $\mean{T}$), we briefly discuss the onset of convection for \eqref{eq:gov_eqs}. The trivial solution of \eqref{eq:gov_eqs} is found by taking $\boldsymbol{u}=0$ and considering a steady state where the temperature is independent of time. The conductive temperature profile, 
\begin{equation}
    \label{eq:Tc}
    T_c = \frac12 z(1-z),
\end{equation}
represents the transport of heat by conduction alone. In the case where there is no rotation ($E=\infty$) for the boundary conditions in \eqref{bc:non_uni}, the system becomes linearly unstable for all $R> 37\,325.2$ \citep{Goluskin2016book}. However, unlike Rayleigh-B\'enard convection when considering \eqref{eq:gov_eqs} in the form
\begin{equation}
    \frac{\textrm{d}\boldsymbol{s}}{\textrm{d}t} + \mathcal{L} \boldsymbol{s} + \mathcal{N}(\boldsymbol{s},\boldsymbol{s}) + \mathcal{P}p = 0 ,
\end{equation}
where $\mathcal{L}$ and $\mathcal{P}$ are linear operators, $\mathcal{N}$ is a bilinear operator and $\boldsymbol{s} = (\boldsymbol{u},T)$ the state vector, the linear operator $\mathcal{L}$ of \eqref{eq:gov_eqs} has a non-zero skew-symmetric component and nonlinear instability can occur at $R = 26\,926.6$, implying that subcritical convection can occur for internally heated convection \citep{straughan2013energy,Goluskin2016book}. For $E < \infty$, rotation stabilises the system to vertical motion and inhibits the onset of convection. For the case of RBC, the effect of rotation on the onset of convection is well quantified \citep{chandrasekhar2013hydrodynamic}. In this section, we demonstrate a result for the effect of rotation on the Rayleigh number for linear instability, $R_L$, in IHC.

\subsection{Linear stability}
\label{sec:linear_stability}

The Rayleigh number up to which the flow is linearly stable is identified by analysing the evolution of perturbations from the linearised system of \eqref{eq:gov_eqs}.  
In the non-rotating case, for selected thermal boundary conditions, the marginally stable states are stationary \citep{Davis1969,herron2003}. When the flow is subject to rotation, the condition for steady rolls, as opposed to oscillatory-in-time rolls, at the onset of convection is unknown. For comparison, in the case of RBC, the onset modes are steady provided $Pr \geq 0.68$, such that assuming a sufficiently large $Pr$ fluid removes the question of oscillatory motion in the analysis we carry out here. The precise structure of the motion at the onset should not affect the asymptotic behaviour of $R_L$ for a small Ekman number. 
It is noteworthy that in the case of the rotating internally heated fluid sphere, the first convective modes are always unsteady \citep{roberts1968thermal,Busse1970,jones2000} and recent evidence from numerical simulations confirms the existence subcritical convection \citep{guervilly2016,kaplan2017}.   

Taking the setup as described in \cref{sec:setup}, we start by decomposing the temperature field into perturbations from the conductive profile $T_c(z)$,
\begin{equation}
    T(\boldsymbol{x},t) = \theta(\boldsymbol{x},t) + T_c(z),
\end{equation}
to obtain the temperature perturbation, $\theta$ equation from \eqref{eq:nondim_energy} and boundary conditions of 
\begin{subequations}
\begin{gather}
    \partial_t \theta + \boldsymbol{u}\cdot\bnabla \theta = \bnabla^2 \theta + T_c' w, \\
    \theta|_{z=\{0,1\}} = 0,
\end{gather}
\end{subequations}
where primes denote derivative with respect to $z$.
Then, we look at the marginally stable stationary states of the linearised system of \eqref{eq:gov_eqs} by considering the $z$ component of the double curl of the momentum equation and the vertical component of the vorticity equation; we thus have
\begin{subequations}
    \label{eq:linear_eqs}
    \begin{align}
        \bnabla^4 w &= E^{-1}\partial_z \zeta - R \bnabla_H^2 \theta \, , \\
        \bnabla^2 \zeta &= - E^{-1} \partial_z w\, , \\
        \bnabla^2 \theta &= T_c' w.
    \end{align}
\end{subequations}
Given horizontal periodicity, we take a Fourier series expansion in the horizontal ($x$ and $y$) directions of the form,
\begin{equation}\label{e:Fourier_f}
    \begin{bmatrix}
    \theta(x,y,z)\\ \boldsymbol{u}(x,y,z) \\ \zeta(x,y,z)
    \end{bmatrix}
    = \sum_{\boldsymbol{k}} 
    \begin{bmatrix}
    \theta_{\boldsymbol{k}}(z)\\ \boldsymbol{u}_{\boldsymbol{k}}(z) \\ \zeta_{\boldsymbol{k}}(z)
    \end{bmatrix}
    \textrm{e}^{i(k_x x + k_y y)}\, ,
\end{equation}
where the sum is over wavevectors $\boldsymbol{k} = (k_x,k_y)$ with magnitude $k = \sqrt{k_x^2 + k_y^2}$ and variables with subscripts $\boldsymbol{k}$ are functions of $z$ only. Then, substituting \eqref{e:Fourier_f} into \eqref{eq:linear_eqs} gives
\begin{subequations}
    \label{eq:linear_gov_f}
    \begin{align}
        \label{eq:ode_1}
        (D^2 - k^2)^2 w_{\boldsymbol{k}} &= E^{-1} \zeta'_{\boldsymbol{k}} + R k^2 \theta_{\boldsymbol{k}} \,, \\
        \label{eq:ode_2}
        (D^2 - k^2) \zeta_{\boldsymbol{k}}  &= - E^{-1}  w'_{\boldsymbol{k}}\, , \\
        \label{eq:ode_3}
        (D^2 - k^2) \theta_{\boldsymbol{k}} &= T_c' w_{\boldsymbol{k}}\, ,
    \end{align}
\end{subequations}
where $D^2 = \frac{\textrm{d}^2}{\textrm{d}z^2}$. 
For no-slip or stress-free boundary conditions,   $\zeta_{\boldsymbol{k}}$ and $\theta_{\boldsymbol{k}}$ can be eliminated from  \eqref{eq:linear_gov_f} to give
\begin{equation}
    (D^2 - k^2)^3 w_{\boldsymbol{k}} + E^{-2} w''_{\boldsymbol{k}} = Rk^2 T_c' w_{\boldsymbol{k}}.
    \label{eq:ODE_RL}
\end{equation}
Here, \eqref{eq:ODE_RL} is an eigenvalue problem and can be solved numerically by fixing $E$ and finding the $R$ at which the first eigenvalue changes sign. 
The equivalent problem for RBC is well documented \citep{chandrasekhar2013hydrodynamic}, however unlike RBC, the ODE in \eqref{eq:ODE_RL} admits solutions which are hypergeometric functions because $T_c$ in \eqref{eq:Tc} is a non-constant function. As such, even in the case of stress-free boundary conditions where $w_{\boldsymbol{k}} = w''_{\boldsymbol{k}} = w''''_{\boldsymbol{k}} = 0$ at both boundaries, \eqref{eq:ODE_RL} becomes complicated to solve. Instead, we consider the asymptotic regime of small $E$ where a simplified form of \eqref{eq:ODE_RL} gives the desired asymptotic relation between $R_L$ and $E$. Following the argument first presented in \cite{chandrasekhar2013hydrodynamic}, we posit that for $R$ close to $R_L$, the wavenumber $k$ tends to infinity so that we retain only terms in $E$, $R$ and the highest power in $k$ and \eqref{eq:ODE_RL} becomes
\begin{equation}
    \label{eq:simp_ODE_LS}
   E^{-2} w''_{\boldsymbol{k}} = k^2(R(\tfrac12 - z ) + k^4) w_{\boldsymbol{k}}.
\end{equation}
Since \eqref{eq:simp_ODE_LS} is of second order, we require only two boundary conditions. However, in the simplest set of boundary conditions of stress-free boundaries where $w_{\boldsymbol{k}}$ and $w_{\boldsymbol{k}}''$ are zero, the problem is over-determined with four boundary conditions. It suffices to take $w_{\boldsymbol{k}}(0)=w''_{\boldsymbol{k}}(0) = 0$, such that we can make the ans\"atz that
\begin{equation}
    \label{eq:ansatz_LS}
    w_{\boldsymbol{k}} = Ai(n -mz) - \frac{Ai(n)}{Bi(n)} Bi(n - m z),
\end{equation}
where $Ai(z)$ and $Bi(z)$ are the Airy functions of the first and second kind.
Substituting \eqref{eq:ansatz_LS} back into \eqref{eq:simp_ODE_LS} gives
\begin{equation}
    E^{-2}m^2(n-m z) = \frac12 k^2 R + k^6 - k^2 R z,
\end{equation}
from which we require the choice that $m = (k^2 R E^2)^\frac13$. Substituting for $m$ and rearranging, we obtain
\begin{equation}
    -\frac12 E^{2/3} k^{2/3} R + n  R^{2/3} - 
E^{2/3} k^{ 14/3 } = 0.
\label{eq:cubic}
\end{equation}%
\begin{figure}
    \centering
    \includegraphics[scale=0.9]{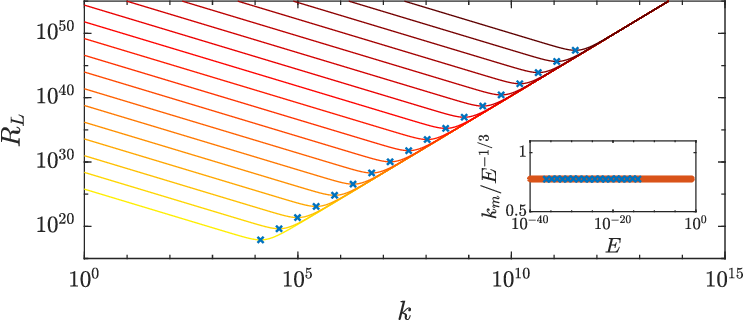}
    \caption{Plot of $R_L$ against $k$ as given by \eqref{eq:Rc_1} for Ekman numbers ranging from $10^{-37}$ ({\color{colorbar17}\solidrule}) to $10^{-13}$ ({\color{colorbar1}\solidrule}). The blue dots represent the minimum $R_L$ in $k$, and the inset demonstrates that the minimum wavenumber $k_m$ varies as $E^{-1/3}$ with a compensated plot. }
    \label{fig:RL}
\end{figure}%
Given that \eqref{eq:cubic} is a cubic equation in $R^{1/3}$, by application of the cubic formula, we find the real root to be, 
\begin{align}
    R_L^{1/3} &= \frac{2 n}{3  E^{ 2/3 } k^{2/3}}
    + \frac{1}{3E^{2/3} k^{2/3}} \left( -8n^3 + 27 E^2 k^6 + 27 E^2 k^3 \sqrt{k^{6} - \frac{16n^3}{27 E^2}} \right)^{\frac13} \nonumber\\
  & + \frac{4n^2}{3 E^{2/3} k^{2/3} } \left( -8n^3 + 27 E^2 k^6 + 27 E^2 k^3 \sqrt{k^{6} - \frac{16n^3}{27 E^2}} \right)^{-\frac13}.
    \label{eq:Rc_1}
\end{align}
Then, we want to find the smallest possible $R_L$ in \eqref{eq:Rc_1} by finding the minimising $k$ by solving $\partial_k R_L^{1/3} = 0$ and substituting back into \eqref{eq:Rc_1} to obtain $R_L$ as a function of only $E$. In \cref{fig:RL} we plot $R_L$ from \eqref{eq:Rc_1} against $k$ for a wide range of $E$ ($10^{-37}$ to $10^{-13}$), highlighting the minimum in $k$ found. The inset in \cref{fig:RL} demonstrates that the $k_m$ varies as $E^{-1/3}$.  
Given \cref{fig:RL}, and noting that in the asymptotic limit of small $E$ we assume large wavenumbers $k$, it is natural in \eqref{eq:Rc_1} to take the minimising wavenumber to be 
\begin{equation}
    k_m^6 = \frac{16n^3}{27E^2},
\end{equation}
such that the terms in the cube roots are real and positive. 
Then, substituting $k_m$ back into \eqref{eq:Rc_1} and the minimal $R_L(n,E)$ is achieved with $n=1$, such that \eqref{eq:Rc_1} simplifies to
\begin{equation}
    R_L  = \frac{(1+2^{2/3}+2^{4/3})^3}{9\cdot 2^{1/3}} E^{-4/3} \sim E^{-4/3}.
    \label{eq:R_E_rel}
\end{equation} 
The asymptotic scaling is equivalent to rotating RBC, unsurprising given the equivalence of the momentum equations, albeit with different prefactors, and highlights the inhibiting effect of rotation on the Rayleigh number for the system to become linearly unstable. 
In the rest of the paper, we will use \eqref{eq:R_E_rel} to arrive at heuristic scaling arguments and later contextualise the bounds proven.

\section{Bounds on the heat flux out of the domain}
\label{sec:Bounds_wT}

In this section, we present proofs of the bounds in \eqref{tab:bound_summary} on $\mathcal{F}_B$ as defined by \eqref{eq:Flux_def}. To obtain a lower bound on $\mathcal{F}_B$, we prove upper bounds on $\wT$ by the background field method in the framework of auxiliary functionals \citep{Arslan2023}. First, in \cref{sec:afm_wT}, we derive the variational problem for finding $U$ where $\wT\leq U$. In \cref{sec:pre_est_wt}, we outline the preliminary choices that are made for the proofs and in \cref{sec:est_bound}, estimate the upper bound on $\wT$. Then, we first prove a bound on $\wT$ valid for large Ekman numbers $E$ in \cref{sec:large_E_wt} by the use of \cref{lem:yan}, followed by a proof valid for small $E$ in \cref{sec:small_E_wT} by using \cref{lem:const}. To provide an overview, \cref{fig:upper_bounds_RIHC_3} illustrates the lower bounds on $\mathcal{F}_B$, omitting the logarithmic corrections for brevity.

\subsection{The auxiliary functional method}
\label{sec:afm_wT}

Here, we outline the main steps in constructing the variational problem to obtain an upper bound on $\wT$. See previous works for a detailed derivation \citep{Arslan2021,Arslan2023}. To prove an upper bound on $\smash{\mean{wT}}$, we employ the auxiliary function method~\citep{Chernyshenko2014a,Fantuzzi2022}. The method relies on the observation that the time derivative of any bounded and differentiable functional $\mathcal{V}\{T(t)\}$ along solutions of the Boussinesq equations~\eqref{eq:gov_eqs} averages to zero over infinite time, so that
\begin{equation}\label{e:af-method}
\mean{wT} = \timeav{\volav{wT} + \tfrac{\rm d}{ {\rm d} t}\mathcal{V}\{T(t)\}  }.
\end{equation}
Two key simplifications follow. The first is that we can estimate \eqref{e:af-method} by the pointwise-in-time maximum along the solutions of the governing equations, and this value is estimated by the maximum it can take over \textit{all} velocity and temperature fields that are periodic in $x$ and $y$, satisfying incompressibility \eqref{eq:continuit}, the boundary conditions \eqref{bc:non_uni} and the maximum principle \cref{lem:min}.

\begin{figure} 
    \centering
    \begin{tikzpicture}[every node/.style={scale=0.95}]
    \draw[->,black,very thick] (-7,0) -- (4.8,0) ;
    \draw[->,black, very thick] (-6.6,-0.4) -- (-6.6,3.5);
    \node at (5,0) {$E$};
    \node at (-7,3.6) {$\mathcal{F}_B$};
    \draw[black,dotted] (-4,0) -- (-4,3.1);
    \draw[black,dotted] (-1,0) -- (-1,3.1);
    \draw[black,dotted] (-6.6,3.1) -- (4.5,3.1);
   \draw[black,dotted] (2,0) -- (2,3.1);
   
    \draw[matlabblue,very thick] plot[smooth,tension=1] coordinates{(-6.6,0) (-4,1.4) (-1,2)};
    \draw[matlabblue,very thick] (-1,2) -- (2,2);
    \draw[matlabgreen,very thick,dashed] (-6.6,1)--(-4,1);
    \draw[matlabblue,very thick,dashdotted] (2,2) -- (4.5,2);
    \node at (-2.5,2.4) {\scriptsize${\color{matlabblue} E^{2/3}R^{-2/3}} $};
     \node at (0.5,2.4) {\scriptsize${\color{matlabblue} R^{-2/3}} $};
     \node at (-5.5,1.4) {\scriptsize${\color{matlabgreen}  R^{-1} } $};
    \node at (3.5,2.4) {\scriptsize${\color{matlabblue} R^{-2/3} } $};
    \node at (3.5,2.8) {\scriptsize${\color{matlabred} R^{-40/29}} $};
    \draw[<->] (-0.8,-0.2) -- (1.8,-0.2) ;
    \draw[<->,matlabgreen] (-6.4,-0.35) -- (-4.2,-0.35) ;
    \draw[<->] (-6.4,-0.2) -- (-1.2,-0.2) ;
    \node at (-5.2,0.5) {$\textrm{A}$};
    \node at (-2.5,0.5) {$\textrm{B}$};
    \node at (0.5,0.5) {$\textrm{C}$};
    \node at (3.5,0.5) {$\textrm{D}$};
    \node at (-5.2,-0.6) { \color{matlabgreen} \tiny $ E \leq R^{-2} $};
    \node at (-6.6,-0.7) {$0$};
    \node at (-7.2,0) {$0$};
    \node at (2.05,-0.22) {\tiny $E_m$};
    \node at (-1,-0.22) {\tiny $E_0$};
    \node at (-7,3.1) {$\frac{1}{2}$};
    \end{tikzpicture}
    \caption[upper bounds RIHC]{ Illustrations of the lower bounds on the flux out of the bottom boundary, $\mathcal{F}_B$, between no-slip and isothermal boundary conditions at infinite $Pr$, the logarithmic corrections in the bounds, \cref{tab:bound_summary}, are suppressed for brevity. The blue plot ({\color{matlabblue}\solidrule}) is the bound derived in \cref{sec:large_E_wt}, valid up to $E =E_m\sim 41.4487$ and $ E_0 = 4.1688$. The bound derived in \cref{sec:small_E_wT} is shown with a green line ({\color{matlabgreen}\dashedrule}). The dashed horizontal line denotes when all heat transport is by conduction and $\mathcal{F}_B=\frac12$. In D ({\color{matlabblue}\dotdashedrule}), we show the bounds for stress-free (top, red) and no-slip (bottom, blue) at zero rotation \citep{Arslan2024}. }
    \label{fig:upper_bounds_RIHC_3}
\end{figure}

We restrict our attention to quadratic functionals taking the form
\begin{equation}\label{e:V-IH1}
\mathcal{V}\{T\}: = \volav{ \frac{\bp}{2} |T|^2 - [\tau(z)+z-1] T },
\end{equation}
that are parametrised by a positive constant $\bp > 0$, referred to as the balance parameter and a piecewise-differentiable function $\tau:[0,1] \to \bR$ with a square-integrable derivative that we call the background temperature field.
Here $\tau(z)$ satisfies
\begin{equation}\label{bc:psi}
\tau(0)=1, \qquad \tau(1)=0.
\end{equation}
Introducing a constant, $U$, and rearranging, \eqref{e:af-method} can be written as,
\begin{equation}
    \wT \leq U - U + \volav{ wT} + \tfrac{\textrm{d}}{\textrm{d}t}\mathcal{V}\{T\}   \leq U,
\end{equation}
where the final inequality holds given that, $ U - \volav{wT} - \tfrac{\textrm{d}}{\textrm{d}t}\mathcal{V}\{T\}  \geq0$, where we can substitute for the Lie derivative of $\mathcal{V}\{T\}$ by using \eqref{eq:nondim_energy}.
However, the minimum principle, \cref{lem:min}, is necessary to obtain a $R$-dependent bound on $\wT$ that approaches $\tfrac12$ from below as $R$ increases. The condition is enforced with a Lagrange multiplier,  $\lambda(z)$, so that the problem statement after computations as outlined in previous work \citep{Arslan2021,Arslan2023} becomes
\begin{align}
    \wT \leq \inf_{U,\beta,\tau(z),\lambda(z)} \{U ~|~ \mathcal{S}\{\boldsymbol{u},T\} \geq 0 \quad \forall(\boldsymbol{u},T) \in  \Tspace_+ \} ,
    \label{eq:opt_prob_g}
\end{align}
where
\begin{equation*}
    \Tspace_+ = \{(\boldsymbol{u},T)| \text{horizontally periodic},\bnabla \cdot \boldsymbol{u}=0, \eqref{bc:non_uni}, \eqref{eq:mom_eqs_diag}, T(\boldsymbol{x})\geq 0 ~~ \text{a.e.} ~~ \boldsymbol{x}\in V \}, 
\end{equation*}
provided $\lambda(z)$ is a non-decreasing function, and
\begin{equation}
    \label{e:S_new}
    \mathcal{S}\{\boldsymbol{u},T\} := \volav{\beta|\nabla T|^2 + \tau'(z) wT + (\beta z - \tau'(z) + \lambda(z))\partial_z T + \tau(z)+ U} - \tfrac12.
\end{equation}
Ensuring the positivity of the quadratic terms in \eqref{e:S_new} is referred to as the \textit{spectral constraint} and is defined as
\begin{equation}
    \label{e:spectral_constraint}
    \volav{\beta|\bnabla T|^2 + \tau'(z) wT } \geq 0\, ,
\end{equation}
where  $w$ and $T$ are related by \eqref{eq:mom_eqs_diag} and subject to the boundary conditions \eqref{bc:non_uni}. As has been previously established \citep{Arslan2023}, provided the spectral constraint is satisfied, then the non-negativity of $\mathcal{S}\{\boldsymbol{u},T\}$ is ensured when $U$ is given by
\begin{equation}
    \label{eq:U}
    \wT \leq U :=\frac12 + \inf_{\substack{ \lambda \in L^{2}(0,1)\\ \lambda~\textrm{non decreasing} \\ \volav{\lambda}=-1  } } \volav{\frac{1}{4\beta} \left|\beta\left(z-\frac12\right) - \tau'(z)+\lambda(z) \right|^2 - \tau(z)  } .
\end{equation}

\subsection{Preliminaries}
\label{sec:pre_est_wt}

To establish a bound on $\wT$, we state the main choices used in the proof that minimise $U(\beta,\tau,\lambda)$ as defined in \eqref{eq:U}. We make the following choice of background temperature field,
\begin{equation}
\label{eq:tau_c}
    \tau(z):=
    \begin{dcases}
        1- \frac{1-a}{\delta}z, & 0 \leq z  \leq \delta, \\
        a, & \delta \leq z \leq 1- \varepsilon, \\
        \frac{a(1-\varepsilon)}{\varepsilon} \left(\frac{1-z}{z}\right), & 1-\varepsilon \leq z \leq 1,
    \end{dcases}
\end{equation}
and set $\lambda(z)$ to be
\begin{equation}
\label{eq:lam}
    \lambda(z):= 
    \begin{dcases}
        -\frac{1-a}{\delta}, & 0 \leq z \leq \delta, \\
        -\frac{a}{1-\delta}, & \delta \leq z \leq 1.
    \end{dcases}
\end{equation}
The piecewise functions $\tau(z)$ and $\lambda(z)$ are quantified by the boundary layer widths $\delta\in(0,\tfrac13)$ and $\varepsilon\in(0,\tfrac13)$, where $\delta \leq \varepsilon$, and parameter $a>0$ that determines the value of $\tau(z)$ in the bulk. 
See  \cref{fig:psi_inf_pr} for a sketch of the functions.

\begin{figure}
	\centering
    \begin{tikzpicture}[every node/.style={scale=0.9}, scale=0.75]
	\draw[->,black,thick] (-12.25,0.5) -- (-5,0.5) node [anchor=north] {$z$};
	\draw[->,black,thick] (-12,0.4) -- (-12,3.8) node [anchor=south] {$\tau(z)$};
	\draw[matlabblue,very thick] (-12,3.3) -- (-11,1.5) ;
    \draw[matlabblue,very thick] (-11,1.5) -- (-6.6,1.5);
    \draw[matlabblue, very thick] plot [smooth, tension = 1] coordinates {(-6.6,1.5) (-6,0.8) (-5.5,0.5)};
	\node[anchor=north] at (-5.5,0.5) {$1$};
	\node[anchor=east] at (-12,3.2) {$1$};
    \draw[dashed] (-11,0.5) node[anchor=north] {$\delta$} -- (-11,1.5);
	\draw[dashed] (-6.6,0.5) node[anchor=north] {$1-\varepsilon$} -- (-6.6,1.5);
	\draw[->,black,thick] (-3.25,3.4) -- (4,3.4) node [anchor=west] {$z$};
	\draw[->,black,thick] (-2.5,0.4) -- (-2.5,3.8) node [anchor=south] {$\lm(z)$};
	\draw[colorbar12,very thick] (-2.5,0.5) -- (-1.5,0.5);
	\draw[colorbar12,very thick] (-1.5,3) -- (3.6,3) ;
	\draw[colorbar12,dotted] (-1.5,3) -- (-1.5,0.5) ;
 \draw[colorbar12,dotted] (3.6,3) -- (3.6,3.4) ;
	\node[anchor=south] at (3.6,3.4) {$1$};
	\node[anchor=south] at (-1.5,3.4) {$\delta$};
	\node[anchor=east] at (-2.5,0.5) {$-\frac{1-a}{\delta} $};
    \end{tikzpicture}
	\caption{Sketches of the functions $\tau(z)$ in~\eqref{eq:tau_c} and $\lambda(z)$ in~\eqref{eq:lam} used to prove \eqref{tab:bound_summary}, where $\delta$ is the boundary layer width at the bottom, $\varepsilon$ the boundary layer width at the top of the domain and $a$ is \cref{eq:a_choice}.}
	\label{fig:psi_inf_pr}
\end{figure}

We will further fix
\begin{equation}
    \label{eq:a_choice}
    a = \frac12 \delta \varepsilon^{1/2},
\end{equation}
and 
\begin{equation}
    \label{eq:beta_choice}
    \beta =  \frac{\volav{|\tau'(z)-\lambda(z)|^2}^{1/2}}{\volav{|z-\tfrac12|^2}^{1/2}} .
\end{equation}

In the following subsections, we prove bounds for different regimes of the Ekman number. We achieve this by using different estimates on the spectral constraint \eqref{e:spectral_constraint}. However, the expression for the upper bound on $\wT$ in \eqref{eq:U} remains the same. Therefore, first, we use our choices of $\tau(z)$ in \eqref{eq:tau_c} and $\lambda(z)$ in \eqref{eq:lam} to estimate \eqref{eq:U}.

\subsection{Estimating the upper bound}
\label{sec:est_bound}

Starting with \eqref{eq:U} an application of the triangle inequality and the choice of $\beta$ in \eqref{eq:beta_choice} gives
\begin{equation}
    \label{eq:U_ineq}
    U \leq \frac12 + \frac{1}{\sqrt{12}} \volav{|\tau'(z) - \lambda(z)|^2}^{1/2} - \volav{\tau}.
\end{equation}
Then, evaluating the sign positive integral with $\tau(z)$ from \eqref{eq:tau_c} and $\lambda(z)$ in \eqref{eq:lam}, gives
\begin{align}
   \volav{  |\tau'(z)-\lambda(z) |^2 } &= \int^{1}_\delta |\tau'(z)-\lambda(z)|^2 \textrm{d}z\nonumber\\ & = \frac{a^2}{(1-\delta)^2}(1-\varepsilon-\delta)  + \frac{a^2}{\varepsilon^2} \int^{1}_{1-\varepsilon} \left(\frac{1-\varepsilon}{z^{2}} - \frac{\varepsilon}{1-\delta}\right)^2 \textrm{d}z\, .
    \label{eq:2norm_exac}
\end{align}
We will require an upper and lower bound on \eqref{eq:2norm_exac}. Starting with a lower bound, given that $\varepsilon\leq \frac13$ and $\delta \leq \frac13$, with $z$ in the range $(1-\varepsilon,1)$, we make the suboptimal but simple estimate that
\begin{equation*}
    \left(\frac{1-\varepsilon}{z^{2}} - \frac{\varepsilon}{1-\delta}\right)^2 \geq \frac19,
\end{equation*}
such that we get
\begin{align}
    \volav{|\tau'(z)-\lambda(z)|^2} \geq \frac{a^2}{\varepsilon^2} \int^{1}_{1-\varepsilon} \left(\frac{1-\varepsilon}{z^{2}} - \frac{\varepsilon}{1-\delta}\right)^2 \textrm{d}z\geq \frac{a^2}{9\varepsilon} .
    \label{eq:lower_U}
\end{align}
For an upper bound on \eqref{eq:2norm_exac}, given that $\varepsilon$ and $\delta$ are positive, bounded above by $\frac13$ and that $\delta\leq \varepsilon$, we use the estimate $(1-\varepsilon-\delta)(1-\delta)^{-2}\leq  \frac34 \varepsilon^{-1} $ and $ z^{-2}\leq  (1-\varepsilon)^{-2}$, to obtain 
\begin{align}
     \volav{|\tau'(z) - \lambda(z)|^2} &    \leq \frac{3a^2}{4\varepsilon} + \frac{a^2}{\varepsilon^2} \int^{1}_{1-\varepsilon} \left(\frac{1-\varepsilon}{z^{2}} - \frac{\varepsilon}{1-\delta}\right)^2 \textrm{d}z \nonumber \\ 
     &\leq \frac{3a^2}{4\varepsilon} + \frac{a^2}{\varepsilon^2} \int^{1}_{1-\varepsilon} \frac{1}{(1-\varepsilon)^2} + \frac{\varepsilon^2}{(1-\delta)^2}  \textrm{d}z \nonumber \\ &\leq \frac{3a^2}{4\varepsilon}  + \frac{a^2}{\varepsilon} \frac{1+\varepsilon^2}{(1-\varepsilon)^2} \leq  \frac{4a^2}{\varepsilon}  . 
     \label{eq:upper_U}
\end{align}
Moving on to the integral of $\tau(z)$ in \eqref{eq:U}, we have that
\begin{equation}
    \label{eq:tau_int}
   \int^{1}_0 \tau(z)~ \textrm{d}z = \frac12 \delta(1-a) - \frac{a}{\varepsilon}(1-\varepsilon)\ln(1-\varepsilon).
\end{equation}
Substituting \eqref{eq:tau_int} and \eqref{eq:upper_U} back into \eqref{eq:U_ineq}, taking $a$ as given by \eqref{eq:a_choice} and $\varepsilon,\delta\leq \frac13$ such that $\delta^2\varepsilon^{1/2} \leq \frac{\sqrt{3}}{9} \delta$, gives
\begin{equation}
    U \leq \frac12 - n  \delta  - \frac{1}{3} \delta \varepsilon^{-1/2} |\ln{(1-\varepsilon)}| \, ,
    \label{eq:U_fa}
\end{equation}
where $n = \frac{18-7\sqrt{3}}{36} $.

\subsection{Large Ekman numbers}
\label{sec:large_E_wt}

To obtain bounds for large $E$ in this subsection, we will use \cref{lem:yan}. The estimates in \cref{lem:yan} are pointwise estimates of the vertical velocity in Fourier space. Therefore, we exploit the horizontal periodicity of $\boldsymbol{u}$ and $T$ and take a Fourier decomposition of $w$ and $T$ in the spectral constraint \eqref{e:spectral_constraint}. Taking that 
\begin{equation}\label{e:Fourier}
    \begin{bmatrix}
    T(x,y,z)\\\boldsymbol{u}(x,y,z)
    \end{bmatrix}
    = \sum_{\boldsymbol{k}} 
    \begin{bmatrix}
    T_{\boldsymbol{k}}(z)\\ \boldsymbol{u}_{\boldsymbol{k}}(z)
    \end{bmatrix}
    \textrm{e}^{i(k_x x + k_y y)}\, ,
\end{equation}
where the sum is over nonzero wavevectors $\boldsymbol{k}=(k_x,k_y)$ for the horizontal periods $L_x$ and $L_y$ and magnitude of each wavevector is 
$k = \sqrt{k_{\smash{x}}^2 + k_{\smash{y}}^2}$. 
Inserting the Fourier expansions~{(\ref{e:Fourier})} into \eqref{e:spectral_constraint} give
\begin{equation}
    \int^{1}_0 \beta |{T}'_{\boldsymbol{k}} |^2 + \beta k^2 |{T}_{\boldsymbol{k}} |^2 + \tau'(z) |{w}_{\boldsymbol{k}} {T}_{\boldsymbol{k}}| \textrm{d}z \geq 0,
    \label{eq:spec_cons_k}
\end{equation}
where the complex conjugate relations of ${w}_{\boldsymbol{k}} = {w}^{*}_{\boldsymbol{k}} $ holds, and $w_{\boldsymbol{k}}$ and $T_{\boldsymbol{k}}$ are subject to the boundary conditions
\begin{subequations}
\begin{gather}
    \label{bc:w_fourier}
    {w}_{\boldsymbol{k}}(0) =   {w}_{\boldsymbol{k}}(1) =   {w}'_{\boldsymbol{k}}(0) =   {w}'_{\boldsymbol{k}}(1) =0, \\
    \label{bc:T_fourier}
     {T}_{\boldsymbol{k}}(0) =   {T}_{\boldsymbol{k}}(1) = 0. 
\end{gather}
\end{subequations}
Based on the boundary conditions, we infer the following two estimates. Given \eqref{bc:w_fourier} applying the fundamental theorem of calculus and H\"olders inequality gives
\begin{equation}
    \label{eq:w_est}
    |{w}_{\boldsymbol{k}}| = \int^{z}_{0} \int^{\sigma}_{0} |\partial^2_\eta {w}_{\boldsymbol{k}}(\eta)| \textrm{d}\eta \textrm{d}\sigma \leq \frac12 z^2 \lVert {w}''_{\boldsymbol{k}} \rVert_{\infty},
\end{equation}
and for ${T}_{\boldsymbol{k}}$, the fundamental theorem of calculus and the Cauchy-Schwarz inequality gives 
\begin{equation}
    \label{eq:T_est}
    |{T}_{\boldsymbol{k}}| = \int^{z}_{0} |\partial_\eta {T}_{\boldsymbol{k}}(\eta)| \textrm{d}\eta  \leq \sqrt{z} \lVert {T}'_{\boldsymbol{k}} \rVert_{2}.
\end{equation}

Next, we substitute \eqref{eq:tau_c} into the sign indefinite term in \eqref{eq:spec_cons_k} to obtain 
\begin{equation}
    \label{eq:sign_indef_int}
    \int^{1}_0 \tau'(z) |{w}_{\boldsymbol{k}} {T}_{\boldsymbol{k}}| \textrm{d}z  = -\frac{1-a}{\delta} \int^{\delta}_0 |{w}_{\boldsymbol{k}} {T}_{\boldsymbol{k}}| \textrm{d}z  - \frac{a(1-\varepsilon)}{\varepsilon} \int^{1}_{1-\varepsilon} z^{-2} |{w}_{\boldsymbol{k}} {T}_{\boldsymbol{k}}| \textrm{d}z . 
\end{equation}
As \cref{lem:yan} contains two estimates for different $\boldsymbol{k}$, we will split the sign indefinite term in half. Then, given that $a\leq1$, use of \eqref{eq:w_est}, the Cauchy-Schwarz inequality, \eqref{eq:w_est_low_k} and \eqref{eq:w_est_high_k} from \cref{lem:yan} gives that
\begin{align}
    \frac{1-a}{ \delta } \int^{\delta}_0 |{w}_{\boldsymbol{k}} {T}_{\boldsymbol{k}} | \textrm{d}z   &\leq \frac{1}{2\delta} \int^{\delta}_0 z^{5/2} \textrm{d}z \lVert {T}'_{\boldsymbol{k}}  \rVert_2 \lVert {w}''_{\boldsymbol{k}} \rVert_\infty \nonumber\\
    &\leq \frac{ \delta^{5/2}}{14} \lVert {T}'_{\boldsymbol{k}} \rVert_2 \lVert {w}''_{\boldsymbol{k}} \rVert_\infty + \frac{\delta^{5/2}}{14} \lVert {T}'_{\boldsymbol{k}} \rVert_2 \lVert {w}''_{\boldsymbol{k}} \rVert_\infty \nonumber \\
    &\leq \frac{c_1}{14} \delta^{5/2} R(1+ \tfrac14 E^{-2})^{1/4} \lVert {T}'_{\boldsymbol{k}} \rVert_2 \lVert {T}_{\boldsymbol{k}} \rVert_2 \nonumber\\ & \quad + \frac{ c_2}{14} \delta^{5/2} \lVert {T}'_{\boldsymbol{k}} \rVert_2  \Big(R \sqrt{k} \lVert {T}_{\boldsymbol{k}} \rVert_2 + R E^{-1} \lVert {T}_{\boldsymbol{k}} \rVert_2   \Big) \, .
    \label{eq:low_int_est_OG}
\end{align}
Taking the term of order $\sqrt{k}$ in \eqref{eq:low_int_est_OG}, we estimate further by noting that from \eqref{eq:T_est} we have a standard Poincar\'e inequality of
 \begin{equation}
    \label{eq:T_norm_est}
     \lVert {T}_{\boldsymbol{k}} \rVert_2 \leq \frac{1}{\sqrt{2}}\lVert {T}'_{\boldsymbol{k}} \rVert_2  ,
 \end{equation}
 such that the use of Youngs' inequality twice gives
 \begin{align}
     \frac{1}{14} \delta^{5/2}   R \sqrt{k} c_2 \lVert {T}'_{\boldsymbol{k}} \rVert_2 \lVert {T}_{\boldsymbol{k}} \rVert_2 &\leq \frac{\beta}{2} k  \lVert {T}_{\boldsymbol{k}} \rVert_2 \lVert {T}'_{\boldsymbol{k}} \rVert_2 + \frac{c_2^2}{392 \beta}   \delta^5 R^2  \lVert {T}_{\boldsymbol{k}} \rVert_2 \lVert {T}'_{\boldsymbol{k}} \rVert_2  \nonumber\\
     &\leq \frac{\beta}{2}k^2 \lVert {T}_{\boldsymbol{k}} \rVert_2^2 + \frac{\beta}{8}\lVert {T}'_{\boldsymbol{k}} \rVert_2^2 + \frac{c_2^2 }{392\sqrt{2}\beta}  \delta^5 R^2 \lVert {T}'_{\boldsymbol{k}} \rVert_2^2.
     \label{eq:est_k_int_1}
\end{align}
Then, substituting \eqref{eq:est_k_int_1} into \eqref{eq:low_int_est_OG}, the integral at the lower boundary becomes
\begin{align}
    \frac{1-a}{\delta}\int^{\delta}_0|{w}_{\boldsymbol{k}}{T}_{\boldsymbol{k}} | \textrm{d}z  \leq & \frac{\beta}{2}k^2 \lVert {T}_{\boldsymbol{k}} \rVert_2^2 + \frac{\beta}{8}\lVert {T}'_{\boldsymbol{k}} \rVert_2^2 + \frac{ c_2^2 }{392 \sqrt{2}\beta}\delta^5 R^2 \lVert {T}'_{\boldsymbol{k}} \rVert_2^2 \nonumber \\   
    &+\frac{ c_1}{14\sqrt{2}}  \delta^{5/2} R (1+ \tfrac14 E^{-2})^{1/4} \lVert {T}'_{\boldsymbol{k}} \rVert_2^2 + \frac{ c_2}{14\sqrt{2}}  \delta^{5/2} R E^{-1} \lVert {T}'_{\boldsymbol{k}} \rVert_2^2.
    \label{eq:low_est}
\end{align}
We realise that for a sufficiently small Ekman number, the term of order $E^{-1}$ is larger than $(1+E^{-2})^{1/4}$ such that if we make the estimate 
\begin{equation}
    \label{eq:upper-E_est}
    c_1(1+\tfrac14 E^{-2})^{1/4} \leq c_2E^{-1} ,
\end{equation}
we get a quadratic form in terms of $E^2$, that places an upper bound on $E$ of  
\begin{equation}
    E\leq E_m =   \frac{1}{2\sqrt{2}}\left( -1 +  \sqrt{1 + 64 (c_2^4 / c_1^4) }  \right)^{1/2} = 41.4487.
    \label{eq:E1}
\end{equation}
Now, \eqref{eq:low_est} becomes, 
\begin{align}
    \frac{1-a}{\delta} \int^{\delta}_0|{w}_{\boldsymbol{k}} {T}_{\boldsymbol{k}} | \textrm{d}z  \leq & \frac{\beta}{2}k^2 \lVert {T}_{\boldsymbol{k}} \rVert_2^2 + \frac{\beta}{8}\lVert {T}'_{\boldsymbol{k}} \rVert_2^2 + \frac{ c_2^2 }{392 \sqrt{2}\beta}\delta^5 R^2 \lVert {T}'_{\boldsymbol{k}} \rVert_2^2    
     + \frac{c_2}{7\sqrt{2}}  \delta^{5/2} R E^{-1} \lVert {T}'_{\boldsymbol{k}} \rVert_2^2.
    \label{eq:low_est2}
\end{align}
Returning to the integral at the upper boundary in \eqref{eq:sign_indef_int}, we apply the same procedure, where \eqref{eq:w_est} and \eqref{eq:T_est} are instead
\begin{equation}
    \label{eq:ests_upper}
    |{w}_{\boldsymbol{k}}| \leq \frac12(1-z)^2 \lVert {w}''_{\boldsymbol{k}} \rVert_\infty, \qquad   |{T}_{\boldsymbol{k}}| \leq \sqrt{1-z} \lVert {T}'_{\boldsymbol{k}} \rVert_2.
\end{equation}
Given $\varepsilon\leq\frac13$ we use that $z^{-2} \leq (1-\varepsilon)^{-2}$ to get
\begin{align}
    \frac{a(1-\varepsilon)}{\varepsilon}\int^{1}_{1-\varepsilon}z^{-2} |{w}_{\boldsymbol{k}}{T}_{\boldsymbol{k}} | \textrm{d}z \leq \, &  \frac{a}{\varepsilon(1-\varepsilon)}\int^{1}_{1-\varepsilon} |{w}_{\boldsymbol{k}}{T}_{\boldsymbol{k}} | \textrm{d}z \leq \frac{3a}{2\varepsilon}\int^{1}_{1-\varepsilon}|{w}_{\boldsymbol{k}}{T}_{\boldsymbol{k}} | \textrm{d}z.
\end{align}
By use of \cref{lem:yan}, along with \eqref{eq:ests_upper}, \eqref{eq:T_norm_est}, Youngs' inequality, and \eqref{eq:upper-E_est}, we can estimate the integral at the upper boundary to obtain
\begin{align}
     \frac{a(1-\varepsilon)}{\varepsilon}\int^{1}_{1-\varepsilon}z^{-2} |{w}_{\boldsymbol{k}}{T}_{\boldsymbol{k}} | \textrm{d}z &\leq  \frac{3a}{2\varepsilon}\int^{1}_{1-\varepsilon}|{w}_{\boldsymbol{k}}{T}_{\boldsymbol{k}} | \textrm{d}z \leq\,  \frac{\beta}{2}k^2 \lVert {T}_{\boldsymbol{k}} \rVert_2^2 + \frac{\beta}{8}\lVert {T}'_{\boldsymbol{k}} \rVert_2^2  \nonumber \\   
    &  + \frac{9c_2^2}{6272\sqrt{2} \beta}   \delta^2 \varepsilon^6 R^2 \lVert {T}'_{\boldsymbol{k}} \rVert_2^2 + \frac{3}{28\sqrt{2}} c_2 \delta \varepsilon^3 R E^{-1} \lVert {T}'_{\boldsymbol{k}} \rVert_2^2 .
    \label{eq:low_est3}
\end{align}
Substituting \eqref{eq:low_est2} and \eqref{eq:low_est3} back into the spectral constraint \eqref{eq:spec_cons_k} gives 
\begin{align}
    \left( \frac{3\beta}{4} - \frac{c_2^2}{392\sqrt{2}}  \frac{\delta^5 R^2}{\beta} - \frac{ c_2}{7\sqrt{2}}  \delta^{5/2} R E^{-1} - \frac{3  c_2}{28\sqrt{2}} \delta \varepsilon^3 R E^{-1} 
    -\frac{9 c_2^2}{6272\sqrt{2} }  \frac{\delta^2 \varepsilon^6 R^2}{\beta } \right) \lVert T'_{\boldsymbol{k}}  \rVert_2^2  \geq 0 .
    \label{eq:spec_c_f}
\end{align}
The spectral condition is satisfied provided the term in the brackets of \eqref{eq:spec_c_f} is non-negative. Note that we have an explicit expression for $\beta$ in \eqref{eq:beta_choice}, which in conjunction with the lower bound in \eqref{eq:lower_U} gives the following lower bound on $\beta$ of
\begin{equation}
    \label{eq:lower_bound_b}
    \beta \geq \frac{\sqrt{3} }{3} \delta.
\end{equation}
After estimating $\beta$ from below with \eqref{eq:lower_bound_b} and making the choice
\begin{equation}
    \label{eq:del_eps_1}
    \delta = \left(\frac{9}{16}\right)^{1/3}  \varepsilon^2,
\end{equation}
the condition for the positivity of \eqref{eq:spec_c_f} becomes, after rearranging,
\begin{equation}
    1 - \frac{c_2^2}{49\sqrt{2}} \delta^3 R^2 - \frac{8 \, c_2}{7\sqrt{6}} \delta^{3/2} R E^{-1} \geq 0 .
    \label{eq:f_cons_spec}
\end{equation}
In \eqref{eq:f_cons_spec}, two possible choices of $\delta=\delta(R,E)$ guarantee the non-negativity of the left-hand side. If the second negative term dominates the first, i.e. 
\begin{equation*}
    \frac{c_2^2}{49\sqrt{2}} \delta^3 R^2 \leq \frac{8 \, c_2}{7\sqrt{6}} \delta^{3/2} R E^{-1},
\end{equation*}
then \eqref{eq:f_cons_spec} becomes
\begin{equation}
\label{eq:delta_1}
    \delta \leq \left(\frac{7\sqrt{6}}{16c_2 }\right)^{2/3} R^{-2/3}E^{2/3}.
\end{equation}
Taking $\delta$ as large as possible in \eqref{eq:delta_1} and substituting back into \eqref{eq:f_cons_spec} implies that $E\leq 8 (\sqrt{2}/3)^{1/2}$. Whereas in the opposite scenario where, 
\begin{equation*}
    \frac{8 \, c_2}{7\sqrt{6}} \delta^{3/2} R E^{-1} \leq \frac{c_2^2}{49\sqrt{2}} \delta^3 R^2 ,
\end{equation*}
then \eqref{eq:f_cons_spec} becomes
\begin{equation}
    \label{eq:delta_2}
    \delta \leq  \left( \frac{49\sqrt{2}}{2c_2^2} \right)^{1/3} R^{-2/3},
\end{equation}
which holds for $E\geq 8(\sqrt{2}/3)^{1/2}$. In summary, the spectral condition holds if the condition in \eqref{eq:f_cons_spec} is satisfied and \eqref{eq:f_cons_spec} is guaranteed when we take $\delta$ as large as possible in \eqref{eq:delta_1} and \eqref{eq:delta_2}. As a result, we have that
\begin{equation}
\label{eq:delta_1f}
\delta =
\begin{dcases}
    \left(\frac{7\sqrt{6}}{16c_2}\right)^{2/3} R^{-2/3}E^{2/3}, & E \leq 8(\sqrt{2}/3)^{1/2} , \\
     \left( \frac{49\sqrt{2}}{2c_2^2} \right)^{1/3} R^{-2/3}, & E \geq 8(\sqrt{2}/3)^{1/2}.
\end{dcases}    
\end{equation}
and by \eqref{eq:del_eps_1} that,
\begin{equation}
\label{eq:eps_1}
\varepsilon =
\begin{dcases}
    \left(\frac{7\sqrt{6}}{12c_2}\right)^{1/3} R^{-1/3}E^{1/3}, & E \leq 8(\sqrt{2}/3)^{1/2}, \\
    \left( \frac{392\sqrt{2}}{9c_2^2} \right)^{1/6} R^{-1/3}, & E \geq 8(\sqrt{2}/3)^{1/2}.
\end{dcases}    
\end{equation}
Therefore, substituting \eqref{eq:delta_1f} and \eqref{eq:eps_1} back into \eqref{eq:U_fa}, along with the fact that $\wT \leq U$ and \eqref{eq:Flux_def} to obtain
\begin{equation}
\label{eq:thm_1}
\mathcal{F}_B \geq 
\begin{dcases}
    d_1 R^{-2/3}E^{2/3} + d_2 R^{-1/2}E^{1/2} | \ln (1- d_3 R^{-1/3}E^{1/3})|  , & E \leq 8(\sqrt{2}/3)^{1/2}, \\
     d_4 R^{-2/3} + d_5 R^{-1/2} | \ln(1-d_6 R^{-1/3})| , &  8(\sqrt{2}/3)^{1/2} \leq E \leq E_m,
\end{dcases}    
\end{equation}
where the constants $d_1$ to $d_6$ are collated in \cref{tab:constants}.
Finally, in \cref{sec:pre_est_wt} we chose that both boundary layer widths are in $(0,\frac13)$, therefore given \eqref{eq:delta_1f} and \eqref{eq:eps_1} the bound obtained in \eqref{eq:thm_1} holds for all $R \geq 0.4715 $.

\subsection{Small Ekman numbers}
\label{sec:small_E_wT}

Next, we demonstrate a proof of the bound on $\mathcal{F}_B$ valid for small $E$ in \eqref{tab:bound_summary}. Here, we use \cref{lem:const} to demonstrate the non-negativity of the spectral constraint \eqref{e:spectral_constraint}. The estimates used in this subsection do not require estimates in Fourier space.

Starting with the spectral constraint in \eqref{e:spectral_constraint}, we start by substituting for $\tau(z)$ from \eqref{eq:tau_c} into the sign-indefinite term and using the estimate $z^{-2} \leq (1-\varepsilon)^{-2}$ at the upper boundary gives 
\begin{equation}
\label{eqa:wt_spec}
    \volav{\tau'(z)wT} \geq - \frac{1-a}{\delta} \left\langle\int^{\delta}_0 wT \textrm{d}z \right\rangle_h - \frac{a}{\varepsilon}(1-\varepsilon)^{-1} \left\langle\int^{1}_{1-\varepsilon} wT \textrm{d}z \right\rangle_h  \, .
\end{equation}
We first consider the integral in \eqref{eqa:wt_spec} near $z=0$ and obtain an estimate on $wT$. Since we require a lower bound on the right-hand side of \eqref{eqa:wt_spec}, we can rearrange the order of integration of the first term on the right-hand side of \eqref{eqa:wt_spec} and estimate the integral from above. Given the boundary conditions \eqref{bc:non_uni}, use of the fundamental theorem of calculus, \eqref{e:z_w_eq} and integration by parts gives 
\begin{align}
    \horav{|wT|} &= \left\langle \left|\int^{z}_0 \partial_s (wT) \textrm{d}s\right| \right\rangle_h = \left\langle \left| \int^{z}_0 T\partial_s w + w \partial_s T \textrm{d}s \right| \right\rangle_h = \left\langle \left|\int^{z}_0 - E T \bnabla^2 \zeta + w \partial_s T \textrm{d}s\right| \right\rangle_h \nonumber\\
    &=  E \horav{|\zeta' T|} + \left\langle \left|\int^{z}_0 E\bnabla \zeta \bnabla T + w \partial_s T \textrm{d}s \right| \right\rangle_h
    .
    \label{eq:wT_est_s}
\end{align}
Then, given the boundary condition on the velocity and temperature in \eqref{bc:non_uni}, we have that
\begin{equation}
    \label{eq:est_simp}
    |w(\cdot,z)| \leq \frac23 z^{3/2}\left(\int^{1}_0 |w''(\cdot,z)|^2 \textrm{d}z\right)^{1/2}, \quad \horav{T^2}^{1/2} \leq \sqrt{z} \volav{|\bnabla T|^2}^{1/2},
\end{equation}
use of which, along with multiple applications of the Cauchy-Schwarz inequality in \eqref{eq:wT_est_s}, and that for $f\in L^2(0,1)$ we have $\horav{|f'|^{2}} \leq \horav{|\bnabla f|^2} \leq \volav{|\bnabla f|^2}$, gives
\begin{align}
    \horav{|wT|} &\leq E \horav{|\zeta'|^2}^{1/2} \horav{|T|^2}^{1/2} + E \volav{|\bnabla\zeta|^2}^{1/2}\volav{|\bnabla T|^2}^{1/2} +\frac{z^2}{3} \left\langle\int^{z}_0 |\partial_s T|^2 \textrm{d}s\right\rangle^{1/2}_h\volav{|w''|^2}^{1/2} \nonumber\\
    &\leq \left(  (\sqrt{z} +1 )E\volav{|\bnabla \zeta|^2}^{1/2}  + \frac13 z^2 \volav{|\bnabla^2 w|^2}^{1/2}\right) \volav{|\bnabla T|^2}^{1/2} \, .
\end{align}
Next, we use \cref{lem:const}, to bound both $\volav{|\bnabla \zeta|^2}$ and $\volav{|\bnabla^2 w|^2}$ from above and given that $T$ is horizontally periodic with Dirichlet boundary conditions at $z=0$ and $1$ we have the standard Poincar\'e inequality $\volav{|T|^2} \leq (1/\pi^2) \volav{|\bnabla T|^2}$, which gives
\begin{equation}
    \label{eq:wT_est_small_E}
    \horav{|wT|} \leq \left(\frac{E\,R}{\sqrt{2}\, \pi}\left(1 + \sqrt{z} \right) + \frac{1}{3\pi}z^2R \right) \volav{|\bnabla T|^2}.
\end{equation}
Then, substituting back into the integral at the boundary, whereby \eqref{eq:a_choice} we have $1-a\leq1$, we get that
\begin{equation}
    \frac{1-a}{\delta} \int^{\delta}_0 \horav{|wT|} \textrm{d}z \leq  \left( \frac{E\, R}{\sqrt{2}\, \pi}\left(1+\frac{2}{3} \delta^{1/2}\right) + \frac{1}{9\pi}\delta^2 R \right) \volav{|\bnabla T|^2}.
    \label{eq:lower_2}
\end{equation}
The same estimates hold at the upper boundary with $a$ given by \eqref{eq:a_choice} and $z$ replaced by $1-z$. Then, since $\varepsilon\leq \frac13$ we can take $1-\varepsilon\leq 1$ to obtain,
\begin{equation}
   \frac12\delta \varepsilon^{-1/2}(1-\varepsilon)^{-1} \int^{1}_{1-\varepsilon} \horav{|wT|} \textrm{d}z \leq  \frac34 \delta\varepsilon^{1/2}\left( \frac{E\,R}{\sqrt{2}\,\pi}\left(1+\frac{2}{3} \varepsilon^{1/2}\right)  + \frac{1}{9\pi}\varepsilon^2 R\right) \volav{|\bnabla T|^2}.
\label{eq:upper_2}
\end{equation}
Substituting \eqref{eq:lower_2} and \eqref{eq:upper_2} back into \eqref{eqa:wt_spec} and then into  \eqref{e:spectral_constraint}, gives after use of the lower bound on $\beta$ from \eqref{eq:lower_bound_b} that 
\begin{equation}
    \left(\frac{\sqrt{3}\delta}{3}  - \frac{E\,R}{\sqrt{2}\pi}\left(1 + \frac{2\delta^{1/2}}{3} \right) - \frac{\delta^2 R}{9\pi} -  \frac{3\delta\varepsilon^{1/2}E\,R}{4\sqrt{2}\pi} \left( 1 + \frac{2\varepsilon^{1/2}}{3} \right) - \frac{\delta\varepsilon^{5/2}R}{12\pi}  \right)\volav{|\bnabla T|^2}\geq 0.
\end{equation}
Since $\varepsilon\leq\frac13$ and $\delta \leq \frac13$, then, we will make estimates, $1+ 2\delta^{1/2}/3 \leq \sqrt{2}\pi/3$ and $ 1+ 2\varepsilon^{1/2}/3 \leq \sqrt{2}\pi/3 $, such that after rearranging the spectral constraint becomes
\begin{equation}
    \sqrt{3}\delta  -  E R  - \frac{\delta^2 R}{3\pi} -   \frac34 \delta\varepsilon^{1/2} E R- \frac{\delta\varepsilon^{5/2}R}{4\pi}  \geq 0.
    \label{eq:cond_small_E}
\end{equation}
In \eqref{eq:cond_small_E}, the first negative term does not contain an explicit $\delta$ dependence, and so we will, at the very least, choose that 
\begin{equation}
   E R \leq  \frac{\sqrt{3}}{2} \delta .
   \label{eq:cond_1_del_wt}
\end{equation}
Using \eqref{eq:cond_1_del_wt} and further making the choice
\begin{equation}
    \label{eq:delt_eps_2}
    \delta = \frac34 \varepsilon^{5/2},
\end{equation}
gives
\begin{equation}
    \frac{\sqrt{3}}{2} - \frac{2}{3\pi}\delta R -  \left(\frac34\right)^{4/5}
 \delta^{1/5} E R  \geq 0.
    \label{eq:cond_small_E_f}
\end{equation}
Similar to the proof of a bound for large Ekman numbers in \cref{sec:large_E_wt}, we have a constraint that we consider in two cases. If in \eqref{eq:cond_small_E_f}, the second negative term dominates the first such that
\begin{equation*}
    \frac{2}{3\pi} \delta R \leq \left(\frac34\right)^{4/5} \delta^{1/5} E R,
    \label{eq:ineq1}
\end{equation*}
then 
\begin{equation}
    \label{eq:delt_no}
    \delta \leq \frac{\sqrt{3}}{36} E^{-5} R^{-5} \,,
\end{equation}
however, if this $\delta$ is to satisfy the spectral constraint then its implication on \eqref{eq:cond_1_del_wt} and \eqref{eq:cond_small_E_f} need to be checked. If we take $\delta$ to be as large as possible in \eqref{eq:delt_no}, then, up to constants, we get from \eqref{eq:cond_1_del_wt} and \eqref{eq:cond_small_E_f} that
\begin{equation*}
    E \gtrsim R^{-4/5} \quad \textrm{and} \quad E \lesssim R^{-1},
\end{equation*}
which leads to a contradiction, and the initial assumption cannot be true. Assuming instead that,
\begin{equation*}
    \label{eq:ineq2}
 \left(\frac34\right)^{4/5}\delta^{1/5} E R \leq    \frac{2}{3\pi} \delta R,
\end{equation*}
gives in \eqref{eq:cond_small_E_f} that
\begin{equation}
    \label{eq:delta_small_E}
    \delta \leq  \frac{3\sqrt{3}\pi}{8} R^{-1}.
\end{equation}
Taking $\delta$ as large as possible in \eqref{eq:delta_small_E}, the constraints from \eqref{eq:cond_1_del_wt} implies that,
\begin{equation}
    E \leq \frac{9\pi}{16}  R^{-2},
    \label{eq:ER_lims}
\end{equation}
Finally, substituting the largest $\delta$ from \eqref{eq:delta_small_E} into \eqref{eq:delt_eps_2} gives
\begin{equation}
    \label{eq:eps_2}
    \varepsilon = \left(\frac{\sqrt{3}\pi}{2}\right)^{2/5}R^{-2/5}.
\end{equation}
Then, taking $\delta$ as large as possible in \eqref{eq:delta_small_E}, \eqref{eq:eps_2}, substituting into \eqref{eq:U_fa}, along with the fact that $\wT \leq U$ and \eqref{eq:Flux_def} we get that, 
\begin{equation}
    \label{eq:thm2}
    \mathcal{F}_B \geq d_7 R^{-1} + d_8 R^{-4/5}|\ln{(1- d_9 R^{-2/5})} |\, , \qquad E \leq  \frac{9\pi}{16} R^{-2},
\end{equation}
where the constants $d_7$ to $d_9$ are collated in \cref{tab:constants}.
Finally, for completeness, we verify that the choice of $\varepsilon,\delta \in (0,\frac13)$ made in \cref{sec:pre_est_wt} is not restrictive. Since $\delta$ is smaller than $\varepsilon$ as expressed in \cref{eq:delt_eps_2}, taking $\delta$ as large as possible in \eqref{eq:delta_small_E} the bound obtained in \eqref{eq:thm2} holds for all $R \geq6.1216$.

\section{Bounds on the mean temperature }
\label{sec:Bounds_T}

In this section, we prove bounds on $\mean{T}$ with the auxiliary functional method by using the same strategy as in \cref{sec:afm_wT}. First, we derive a variational problem for obtaining a lower bound on $\mean{T}$, with a different quadratic auxiliary functional to \eqref{e:V-IH1}. Then, given our choice of background profile $\phi(z)$, we can estimate the lower bound $L$ on $\mean{T}$ in terms of the parameters of $\phi(z)$. We then prove bounds for large and small $E$ numbers by \cref{lem:yan} and \cref{lem:const}. The proofs in this section are algebraically  lighter than the proof for a bound on $\wT$, primarily due to two reasons. The first is that the minimum principle on $T$, \cref{lem:min}, is not required, and the second, the balance parameters do not have a $R$ dependence.  
The lower bounds we prove on $\mean{T}$, including those already known, are illustrated in \cref{fig:lower_bounds_RIHC_3}.
\begin{figure}
    \centering
    \begin{tikzpicture}[every node/.style={scale=0.95}]
    \draw[->,black,very thick] (-7,0) -- (4.8,0) ;
    \draw[->,black, very thick] (-6.6,-0.4) -- (-6.6,3.5);
    \node at (5,0) {$E$};
    \node at (-7,3.5) {$\mean{T}$};
    \draw[black,dotted] (-3.5,0) -- (-3.5,3);
    \draw[black,dotted] (-0.5,0) -- (-0.5,3);
    \draw[black,dotted] (-6.6,3) -- (4.5,3);
   \draw[black,dotted] (2.5,0) -- (2.5,3);
   
    \draw[matlabblue,very thick] plot[smooth,tension=1] coordinates{(-6.6,0) (-4.2,1.5) (-0.5,2)};
    \draw[matlabblue,very thick] (-0.5,2) -- (2.5,2);
    \draw[matlabgreen,very thick,dashed] plot[smooth,tension=1] coordinates{(-3.5,1.5) (-1,0.5) (3.5,0.1)};
    \draw[matlabgreen,very thick,dashed] (-6.6,1.5) -- (-3.5,1.5);
    \draw[matlabblue,very thick,dashdotted] (2.5,2.3) -- (4.5,2.3);
    \node at (-2,2.4) {\scriptsize${\color{matlabblue} E^{2/7}R^{-2/7}} $};
     \node at (1,2.4) {\scriptsize${\color{matlabblue} R^{-2/7}} $};
     \node at (-5,1.8) {\scriptsize${\color{matlabgreen} R^{-1/3}} $};
     \node at (-0.5,1) {\scriptsize${\color{matlabgreen} E^{-1} R^{-1} } $};
    \node at (3.5,2.5) {\scriptsize${\color{matlabblue} R^{-1/4}\ln{R}^{-1/4}} $};
    \node at (3.5,2.85) {\scriptsize${\color{matlabred} R^{-5/17}} $};
    \draw[<->] (-6.4,-0.2) -- (-0.6,-0.2) ;
    \draw[<->] (-0.3,-0.2) -- (2.3,-0.2) ;
    \draw[<->,matlabgreen] (-6.4,-0.4) -- (-3.8,-0.4) ;
    \draw[<->,matlabgreen] (-3.3,-0.4) -- (3.6,-0.4) ;
    \node at (-5,0.4) {$\textrm{A}$};
    \node at (-2,0.4) {$\textrm{B}$};
    \node at (1,0.4) {$\textrm{C}$};
    \node at (3.8,0.4) {$\textrm{D}$};
    \node at (-4.8,-0.6) {\color{matlabgreen}\tiny $ E \leq R^{-2/3} $};
    \node at (-2,-0.6) {\color{matlabgreen}\tiny $R^{-2/3} \leq E  $};

    \node at (-0.45,-0.25) {\tiny $E_1$};
    \node at (2.55,-0.25) {\tiny $E_m$};
    \node at (-6.6,-0.7) {$0$};
    \node at (-7.2,0) {$0$};
    \node at (-7,3) {$\frac{1}{12}$};
    \end{tikzpicture}
    \caption[upper bounds RIHC]{ Illustration of the lower bounds on the mean temperature between no-slip and isothermal boundary conditions at infinite $Pr$. The blue plot ({\color{matlabblue}\solidrule}) is the bound derived in \cref{sec:large_E_T}, valid up to $E_m \sim 41.4487$ and $E_1=8$. The bounds derived in \cref{sec:small_E_T} are shown with a green plot ({\color{matlabgreen}\dashedrule}). The dashed horizontal line denotes when all heat transport is by conduction and $\mean{T}=\frac{1}{12}$. In D ({\color{matlabblue}\dotdashedrule}), we show the previously known best bound for stress-free (top, red) \citep{Whitehead2012} and no-slip (bottom, blue) \citep{Whitehead2011} at zero rotation. }
    \label{fig:lower_bounds_RIHC_3}
\end{figure}

\subsection{The auxiliary functional method}
\label{sec:afm_T}

By the auxiliary functional method, we obtain an explicit variational problem that we solve to obtain a bound on $\mean{T}$. The following derivation, albeit in the language of the classic background field method approach, appears in previous papers \citep{Whitehead2011,Whitehead2012}. Here, we only present an outline within the auxiliary functional framework. 

Starting with the quadratic auxiliary functional
\begin{equation}
    \label{eq:V_T}
    \mathcal{V}\{T\} = \volav{-\frac{1}{2}|T|^2 + 2\phi(z) T },
\end{equation}
where $\phi(z)$ is the background temperature field subject to the boundary conditions
\begin{equation}
\phi(0)=\phi(1)=0.    
\end{equation}
Then, provided $\mathcal{V}\{T(t)\}$ is bounded along solutions of \eqref{eq:gov_eqs}, the time derivative of the long-time average of $\mathcal{V}\{T\}$ is zero, such that we can write
\begin{equation}
    \mean{T}=L - \left(L - \overline{\volav{T} + \tfrac{\textrm{d}}{\textrm{d}t}\mathcal{V}\{T\}} \right) \geq L ,
\end{equation}
where the inequality comes from assuming that, $L-\overline{ \volav{T} + \tfrac{\textrm{d}}{\textrm{d}t}\mathcal{V}\{T\} } \leq L-\volav{T} - \tfrac{\textrm{d}}{\textrm{d}t}\mathcal{V}\{T\} \leq  0 $. Noting that we again bound the terms in the long-time integral by the pointwise in time maximum over all periodic $\boldsymbol{u}$ and $T$, where $\bnabla \cdot \boldsymbol{u}=0$, subject to the boundary conditions \eqref{bc:non_uni}. Hence, after substituting for the Lie derivative of $\mathcal{V}\{T\}$ by use of \eqref{eq:nondim_energy} and rearranging, we have, after appropriate manipulations, the following variational problem
\begin{equation}
    \label{eq:opt_prob_t_r}
    \mean{T} \geq  \inf_{L, \phi(z)} 
    \Big\{
    L \;\vert\;
    \mathcal{S}\{\boldsymbol{u},T\}  \geq 0 \quad \forall (\boldsymbol{u} ,T)\in\mathcal{H}_+
    \Big\},
\end{equation}
where
\begin{equation}
    \label{eq:S_T}
    \mathcal{S}\{\boldsymbol{u},T\} := \volav{|\bnabla T|^2 + \phi'(z)wT - \phi'(z)\partial_z T + \phi(z) - L }.
\end{equation}

In this case, the spectral constraint is given by 
\begin{equation}
    \label{eq:spec_cons_2}
    \volav{|\bnabla T|^2 + 2\phi'(z)wT}\geq 0.
\end{equation}
Then, by optimising the linear terms in \eqref{eq:S_T} the explicit expression for $L$ is given 
\begin{align}
    \label{eq:lower_bound_T}
    L :=&~ 2\volav{\phi(z)}  - \volav{|\phi'(z)|^2} .
\end{align}

\subsection{Preliminaries}
\label{sec:prelim_T}
To establish a lower bound on $\mean{T}$, we take the following class of background fields 
\begin{equation}
    \label{eq:T_phi}
    \phi(z):=
    \begin{dcases}
        \tfrac12(1-\delta)z, & 0 \leq z \leq \delta, \\
        \tfrac12\delta(1-\delta) ,& \delta \leq z \leq 1- \delta,\\
        \tfrac12(1-\delta)(1-z), & 1-\delta \leq z \leq 1.
    \end{dcases}
\end{equation}
The background field is entirely determined by $\delta\in(0,\frac13)$, the width of the boundary layers at the top and bottom of the domain, sketched in \cref{fig:psi_inf_pr_T}. In contrast to \cref{sec:Bounds_wT}, no advantage follows from background fields with boundary layers of different widths.   
\begin{figure}
	\centering
    \begin{tikzpicture}[every node/.style={scale=0.9}, scale=0.75]
	\draw[->,black,thick] (-12.25,0.5) -- (-5,0.5) node [anchor=north] {$z$};
	\draw[->,black,thick] (-12,0.4) -- (-12,3.8) node [anchor=south] {$\phi(z)$};
    \draw[matlabblue,very thick]  (-12,0.5) -- (-11,3) -- (-6.6,3) -- (-5.6,0.5);
	\node[anchor=north] at (-5.6,0.5) {$1$};
    \draw[dashed] (-11,0.5) node[anchor=north] {$\delta$} -- (-11,3);
	\draw[dashed] (-6.6,0.5) node[anchor=north] {$1-\delta$} -- (-6.6,3);
    \draw[dashed] (-12,3) node[anchor=east] {$ \frac12 \delta(1-\delta)$} -- (-11,3);
    \end{tikzpicture}
	\caption{Sketches of the functions $\phi(z)$ in~\eqref{eq:T_phi} used to prove \eqref{tab:bound_summary}, where $\delta$ is the boundary layer widths.}
	\label{fig:psi_inf_pr_T}
\end{figure}

The lower bounds on $\mean{T}$ for different Ekman numbers rely on different estimates of the velocity in the spectral constraint \eqref{eq:spec_cons_2} given the diagnostic equations \eqref{eq:mom_eqs_diag}. Given the choice of background profile \eqref{eq:T_phi}, the expression for the lower bound $L=L(\delta)$ is unchanged based on the estimates we use to demonstrate the non-negativity of the spectral constraint for different Ekman numbers. So, we now estimate $L$ in \eqref{eq:lower_bound_T} given \eqref{eq:T_phi}. Substituting for $\phi(z)$ from \eqref{eq:T_phi}, we get
\begin{equation}
    L  = \frac12 \delta (1-\delta)^2.
\end{equation}
Since $\delta\leq \frac13$, then the estimate $1-\delta \geq \frac23$, gives
\begin{equation}
    \label{eq:L_T}
    L \geq \frac29 \delta \, .
\end{equation}

\subsection{Large Ekman numbers}
\label{sec:large_E_T}

We can now proceed to establish a lower bound on $\mean{T}$ by establishing the conditions under which the spectral constraint of \eqref{eq:opt_prob_t_r} is satisfied, given the choice of background temperature field \eqref{eq:T_phi} and by use of \cref{lem:yan}. The estimates in \cref{lem:yan} apply to \eqref{eq:mom_eqs_diag} in Fourier space so we exploit horizontal periodicity and substitute \eqref{e:Fourier} into \eqref{eq:spec_cons_2} such that the spectral constraint in Fourier space is 
\begin{equation}
    \int^{1}_0 |{T}'_{\boldsymbol{k}}|^2 + k^2|{T}_{\boldsymbol{k}}|^2 + 2\phi'(z)\textrm{Re}\{  {w}_{\boldsymbol{k}} {T}_{\boldsymbol{k}}^*\} \textrm{d}z \geq 0.
    \label{eq:spec2_fourier}
\end{equation}
Substituting for $\phi(z)$ and using the fact that $\delta\leq \frac13$ gives $-(1-\delta) \geq -1$, the sign-indefinite of \eqref{eq:spec2_fourier} is 
\begin{align}
    \label{eq:sign_indef_T}
    2\int^{1}_0 \phi'(z) \textrm{Re}\{  {w}_{\boldsymbol{k}} {T}_{\boldsymbol{k}}^* \} \textrm{d}z  &\geq - 2\int^{1}_0 |\phi'(z) {w}_{\boldsymbol{k}} {T}_{\boldsymbol{k}}| \textrm{d}z \nonumber \\ 
    &\geq  -  \int^{\delta}_0 | {w}_{\boldsymbol{k}} {T}_{\boldsymbol{k}}| \textrm{d}z -  \int^{1}_{1-\delta} | {w}_{\boldsymbol{k}} {T}_{\boldsymbol{k}}| \textrm{d}z\, .
\end{align}
Due to the symmetry in the boundary conditions and $\phi(z)$, we only demonstrate estimates on the integral at the lower boundary since that at the top gives an identical contribution. Using the estimates of \eqref{eq:w_est} and \eqref{eq:T_est} on $w_{\boldsymbol{k}}$ and $T_{\boldsymbol{k}}$, \cref{lem:yan} and Youngs' inequality, we get
\begin{align}
    \int^{\delta}_0 |{w}_{\boldsymbol{k}} {T}_{\boldsymbol{k}} | \textrm{d}z   &\leq \frac{1}{2} \int^{\delta}_0 z^{5/2}  \textrm{d}z \lVert {T}'_{\boldsymbol{k}} \rVert_2 \lVert {w}''_{\boldsymbol{k}} \rVert_\infty \nonumber\\
    &= \frac{1}{14}  \delta^{7/2} \lVert {T}'_{\boldsymbol{k}} \rVert_2 \lVert {w}''_{\boldsymbol{k}} \rVert_\infty + \frac{1}{14}  \delta^{7/2} \lVert {T}'_{\boldsymbol{k}} \rVert_2 \lVert {w}''_{\boldsymbol{k}} \rVert_\infty  \nonumber\\
    &\leq \frac{ \delta^{7/2}}{14} c_1 R(1+ \tfrac14 E^{-2})^{1/4} \lVert {T}_{\boldsymbol{k}}' \rVert_2 \lVert {T}_{\boldsymbol{k}} \rVert_2 \nonumber\\ & \quad + \frac{\delta^{7/2}}{14}  \lVert {T}_{\boldsymbol{k}}' \rVert_2 c_2 \big(R \sqrt{k} \lVert {T}_{\boldsymbol{k}} \rVert_2 + R E^{-1} \lVert {T}_{\boldsymbol{k}} \rVert_2   \big) .
\end{align}
 Taking the term of order $\sqrt{k}$, estimating by use of Youngs' inequality twice and \eqref{eq:T_norm_est} gives
 \begin{align}
     \frac{1}{14} \delta^{7/2}   R \sqrt{k} c_2 \lVert {T}'_{\boldsymbol{k}} \rVert_2 \lVert {T}_{\boldsymbol{k}} \rVert_2 &\leq \frac{\sqrt{2}}{2} k  \lVert {T}_{\boldsymbol{k}} \rVert_2 \lVert {T}'_{\boldsymbol{k}} \rVert_2 +  \frac{c_2^2}{392\sqrt{2}}  \delta^7 R^2   \lVert {T}_{\boldsymbol{k}} \rVert_2 \lVert {T}'_{\boldsymbol{k}} \rVert_2  \nonumber\\ 
     &\leq \frac{1}{2}k^2 \lVert {T}_{\boldsymbol{k}} \rVert_2^2 + \frac{1}{4}\lVert {T}'_{\boldsymbol{k}} \rVert_2^2 + \frac{ c_2^2}{784} \delta^7 R^2 \lVert {T}'_{\boldsymbol{k}} \rVert_2^2.
\end{align}
Therefore, the integral from 0 to $\delta$ simplifies to
\begin{align}
  \int^{\delta}_0|{w}_{\boldsymbol{k}}{T}_{\boldsymbol{k}} | \textrm{d}z  \leq &~ \frac12 k^2 \lVert {T}_{\boldsymbol{k}} \rVert_2^2 + \frac{1}{4}\lVert {T}'_{\boldsymbol{k}} \rVert_2^2 + \frac{ c_2^2}{784}  \delta^7 R^2 \lVert {T}'_{\boldsymbol{k}} \rVert_2^2 \nonumber \\   
    &+\frac{\sqrt{2}}{28}  c_1 \delta^{7/2} R (1+ \tfrac14 E^{-2})^{1/4} \lVert {T}'_{\boldsymbol{k}} \rVert_2^2 + \frac{\sqrt{2}}{28} c_2 \delta^{7/2} R E^{-1} \lVert {T}'_{\boldsymbol{k}} \rVert_2^2.
\end{align}
By equivalent estimates at the upper boundary, we get the same upper bound such that \eqref{eq:sign_indef_T} becomes
\begin{align}
    2\int^{1}_0 \phi'(z) \textrm{Re}\{{w}_{\boldsymbol{k}} {T}_{\boldsymbol{k}}^* \}\textrm{d}z \geq & - k^2 \lVert {T}_{\boldsymbol{k}} \rVert_2^2 - \frac{1}{2} \lVert {T}'_{\boldsymbol{k}}\rVert_2^2 - \Bigg(\frac{c_2^2 }{392} \delta^7 R^2  \nonumber \\
    &+\frac{\sqrt{2}}{14}  c_1 \delta^{7/2} R (1+ \tfrac14 E^{-2})^{1/4}  + \frac{\sqrt{2}}{14} c_2 \delta^{7/2} R E^{-1} \Bigg) \lVert {T}'_{\boldsymbol{k}} \rVert_2^2.
    \label{eq:sign_indef_T2} 
\end{align}
Finally, we use the estimate in \eqref{eq:upper-E_est}, which places an upper limit on $E=E_m$ in \eqref{eq:E1}, for which the bounds are valid. Then, substituting \eqref{eq:sign_indef_T2} into \eqref{eq:spec2_fourier} with \eqref{eq:upper-E_est}, the spectral constraint becomes
\begin{equation}
    \label{eq:spec_con_T_f}
   \frac{1}{2} - \frac{c_2^2}{392}  \delta^7 R^2 - \frac{ \sqrt{2}}{7} c_2 \delta^{7/2} R E^{-1}   \geq 0.
\end{equation}
In \eqref{eq:spec_con_T_f}, two possible choices of $\delta=\delta(R,E)$  guarantee the non-negativity of the left-hand side. If, 
\begin{equation*}
    \frac{c_2^2}{392}  \delta^7 R^2 \leq \frac{ \sqrt{2}}{7} c_2 \delta^{7/2} R E^{-1} ,
\end{equation*}
then from we \eqref{eq:spec_con_T_f} we have that
\begin{equation}
    \label{eq:delta_T1}
    \delta \leq \left( \frac{7\sqrt{2}}{8c_2}\right)^{2/7} R^{-2/7} E^{2/7}.
\end{equation}
Taking \eqref{eq:delta_T1} as large as possible, substituting back into \eqref{eq:spec_con_T_f} and rearranging, we find that the spectral constraint holds when $E \leq 8$. In the opposite scenario where,
\begin{equation*}
     \frac{ \sqrt{2}}{7} c_2 \delta^{7/2} R E^{-1}  \leq \frac{c_2^2}{392}  \delta^7 R^2,
\end{equation*}
substituted back into \eqref{eq:spec_con_T_f} gives that
\begin{equation}
    \delta \leq \left( \frac{196}{c_2^2} \right)^{1/7} R^{-2/7},
\end{equation}
which holds for $E\geq8$.
In summary, we have that
\begin{equation}
    \label{eq:delta_3}
    \delta =
    \begin{dcases}
        \left( \frac{7\sqrt{2}}{8c_2}\right)^{2/7} R^{-2/7} E^{2/7}, & E\leq 8,\\
        \left( \frac{196}{c_2^2} \right)^{1/7} R^{-2/7},& E\geq 8,
    \end{dcases}
\end{equation}
which we can finally substitute in to \eqref{eq:L_T} remembering that $\mean{T}\geq L$ to obtain that
\begin{equation}
    \label{eq:thm3}
    \mean{T} \geq 
    \begin{dcases}
        d_{10} R^{-2/7} E^{2/7}, & E\leq 8,\\
        d_{11} R^{-2/7},& 8 \leq E \leq E_m ,
    \end{dcases}
\end{equation}
where $d_{10}$ and $d_{11} $ are in \cref{tab:constants}.
A final check is on the validity of the bounds given choices made in \cref{sec:prelim_T} that $\delta$ is in $(0,\frac13)$. Given \eqref{eq:delta_3} the bound obtained in \eqref{eq:thm3} holds for all $R \geq 7.1363$ when $E \leq 8$ and $R\geq 10.0922$ when $E \geq 8$.

\subsection{Small Ekman numbers}
\label{sec:small_E_T}

Next, we move on to the proof of the lower bound on $\mean{T}$ in \eqref{tab:bound_summary} valid for a small $E$. Here, we will use \cref{lem:const}, which does not require estimates in Fourier space. Starting with the spectral constraint \eqref{eq:spec_cons_2} and substituting for $\phi(z)$ from \eqref{eq:T_phi}, the sign-indefinite term becomes
\begin{equation}
    \label{eq:sign_indef_T_small}
    2\volav{\phi'(z) w T} \geq -2 \volav{|\phi'(z) wT|} \geq -\left\langle \int^{\delta}_0|wT| \textrm{d}z \right\rangle_h - \left\langle \int^{1}_{1-\delta}|wT| \textrm{d}z \right\rangle_h.
\end{equation}
In \cref{sec:small_E_wT}, we established the estimate \eqref{eq:wT_est_small_E} and can directly substitute into the integral at the lower boundary in \eqref{eq:sign_indef_T_small} to obtain 
\begin{equation}
    \left\langle \int^{\delta}_0|wT| \textrm{d}z \right\rangle_h \leq \left[\frac{E\,R}{\sqrt{2}\pi}\left(\delta +\frac{  2\delta^{3/2}}{3}\right) + \frac{1}{9\pi}\delta^3 R \right] \volav{|\bnabla T|^2}.
\end{equation}
The integral at the upper boundary gives an identical estimate such that the spectral constraint of \eqref{eq:spec_cons_2} becomes
\begin{align}
    \volav{|\bnabla T|^2 + 2\phi'(z)wT} &\geq \volav{|\bnabla T|^2 - 2|\phi'(z)wT| } \nonumber \\ 
    &\geq \volav{|\bnabla T|^2} - 2\left[\frac{E\,R}{\sqrt{2}\pi}\left(\delta +\frac{  2\delta^{3/2}}{3}\right) + \frac{1}{9\pi}\delta^3 R \right] \volav{|\bnabla T|^2} \geq 0.
\end{align}
Therefore, the spectral constraint becomes the condition
\begin{equation}
    1 -  \frac{2 E\,R}{\sqrt{2}\pi}\left(\delta +\frac{  2 \delta^{3/2}} {3 }\right) - \frac{1}{9\pi}\delta^3 R \geq 0 .
\end{equation}
Given that $\delta \leq \frac13$, we use the estimate that $1+2\delta^{1/2}/3 \leq \frac{\sqrt{2}\pi}{3}$ to get the condition,
\begin{equation}
    \label{eq:spec_cond_final_T}
    1 - \frac23 E R \delta - \frac{1}{9\pi} \delta^3 R \geq 0.
\end{equation}
Once more, the condition admits $\delta=\delta(R,E)$ valid for two cases. If we take
\begin{equation*}
    \frac23 E R \delta \leq \frac{1}{9\pi} \delta^3 R,
\end{equation*}
then \eqref{eq:spec_cond_final_T} is non-negative if
\begin{equation}
    \delta \leq \left( \frac{9\pi}{2} \right)^{1/3} R^{-1/3},
\end{equation}
where when taking $\delta$ as large as possible the choice holds when $E \leq \frac{3}{4}\left(\frac{2}{9\pi}\right)^{1/3}R^{-2/3}$. Whereas in the case where
\begin{equation*}
    \frac{1}{9\pi} \delta^3 R \leq  \frac23 E R \delta,
\end{equation*}
then
\begin{equation}
    \delta \leq  \frac{3}{4} R^{-1} E^{-1},
\end{equation}
for all $E \geq  \frac{3}{4}\left(\frac{2}{9\pi}\right)^{1/3}R^{-2/3}$. Therefore, the spectral constraint is satisfied if
\begin{equation}
    \label{eq:delta_4}
    \delta = 
    \begin{dcases}
        \left( \frac{9\pi}{8} \right)^{1/3} R^{-1/3}, & E \leq \frac{3}{4}\left(\frac{2}{9\pi}\right)^{1/3}R^{-2/3}, \\
        \frac{3}{4} R^{-1} E^{-1}, & E \geq \frac{3}{4}\left(\frac{2}{9\pi}\right)^{1/3}R^{-2/3}.
    \end{dcases}
\end{equation}
Substituting into \eqref{eq:L_T} and remembering further that $\mean{T}\geq L$, gives
\begin{equation}
    \label{eq:thm4}
    \mean{T} \geq
    \begin{dcases}
        d_{12} R^{-1/3}, & E \leq \frac{3}{4}\left(\frac{2}{9\pi}\right)^{1/3}R^{-2/3}, \\
        d_{13} R^{-1}E^{-1}, & E \geq \frac{3}{4}\left(\frac{2}{9\pi}\right)^{1/3}R^{-2/3},
    \end{dcases}
\end{equation}
where $d_{12}$ and $d_{13}$ are in \cref{tab:constants}.
Finally, for completeness, we verify that the choice of $\delta \in (0,\frac13)$ made in \cref{sec:prelim_T} is not restrictive. Given \eqref{eq:delta_4} the bound obtained in \eqref{eq:thm4} holds for all $R \geq 95.4259$ when $E \leq \frac{3}{4}\left(\frac{2}{9\pi}\right)^{1/3}R^{-2/3}$ and for all $R\geq1$ otherwise.

\section{Discussion}
\label{sec:conclusion}

\subsection{Regions of validity of the bounds}

\begin{figure}
    \centering
    \hspace{-1cm}
    \includegraphics[scale=0.95]{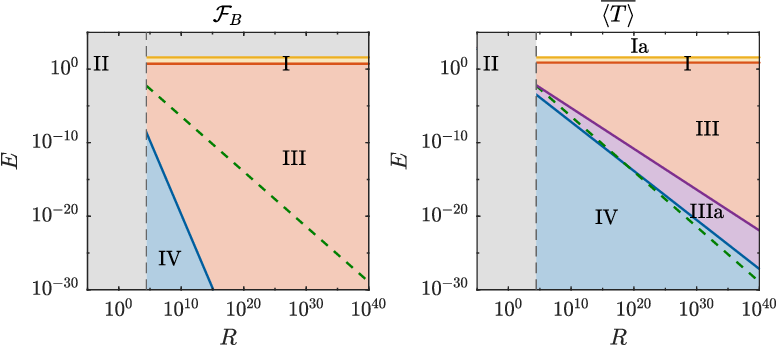}
    \begin{tikzpicture}[overlay]
        \node at (-12.1,4.95) {\textit{(a)}};
        \node at (-5.8,4.95) {\textit{(b)}};
    \end{tikzpicture}
    \caption{The Rayleigh and Ekman number diagrams for the bounds \eqref{eq:thm_1}, \eqref{eq:thm2}, \eqref{eq:thm3} and \eqref{eq:thm4}. Grey areas denote regions where the bounds proven in this work do not hold. We plot the regions of validity for the bounds on $\mathcal{F}_B$ in \textit{(a)} and for $\mean{T}$ in \textit{(b)}. The blue solid lines ({\color{matlabblue}\solidrule}) corresponds to $E= 1.7671 \,R^{-2}$ in \textit{(a)} and $E=0.3102\, R^{-2/3}$ in \textit{(b)}, with the blue area denoting region IV. The horizontal red solid lines ({\color{matlabred}\solidrule}) are $E_0 = 5.4927$ in \textit{(a)} and $E_1 = 8$ in \textit{(b)} and the red area is region III, where for \textit{(b)} IIIa is the purple zone with the solid purple line ({\color{mypurple}\solidrule}) corresponding to $E = 1.9273\, R^{-5/9}$. In both figures, the horizontal yellow lines ({\color{colorbar2}\solidrule}) are $E_m = 41.4487$ with the yellow zone corresponding to region I. The vertical dashed lines are $R_L=26926.6$ and the green dashed lines ({\color{matlabgreen}\dashedrule }) the asymptotic result of \eqref{eq:R_E_rel}. The four main regions are labelled I to IV.}
    \label{fig:bounds_regimes}
\end{figure}

Owing to the use of two different estimates (\cref{lem:yan} and \cref{lem:const}), we need to consider which bounds dominate in the various regions created by the constraints on the bounds. 
To this effect, first, we plot the regions in which the bounds on $\mathcal{F}_B$ and $\mean{T}$ overlap in \cref{fig:bounds_regimes}. For ease of understanding, we split the space of $E$ and $R$ into four main regions. Region I, where $R$ and $E$ are large, corresponds to a slowly rotating buoyancy-dominated flow. Region II is the solid body rotation of the fluid since $R$ and $E$ are small. Region III, where $R$ is large and $E$ small, contains a transition from buoyancy- to rotation-dominated convection provided $R>R_L$ where $R_L$ is the Rayleigh number above which the flow is linearly unstable. At the same time, region IV, where $E$ is small but $R$ cannot get too large, is for rotation-dominated flows. 

In \cref{fig:bounds_regimes}\textit{(a)}, the blue solid line shows $E = 1.7671\, R^{-2}$, the red line is the constant $E_0= 5.4927$, whereas in \cref{fig:bounds_regimes}\textit{(b)} the blue line shows $E = 0.3102 R^{-2/3}$, the purple line $E = 1.9273 R^{-5/9}$ and the red line is the constant $E_1 = 8$. In both \textit{(a)} and \textit{(b)}, the yellow line is the constant $E_m = 41.4487$ from \eqref{eq:E1}, the dotted vertical line, $26926.6$ and the dashed green line is the asymptotic result of \eqref{eq:R_E_rel}.

Starting with \cref{fig:bounds_regimes}\textit{(a)}, the bounds of \eqref{eq:thm_1} and \eqref{eq:thm2} split the diagram into four and we evaluate the best bound in each region. For region I,  where $E\gtrsim  R^{-2}$ and $E_m \geq E\geq E_0$, the only valid bound is $d_4 R^{-2/3} + d_5 R^{-1/2}|\ln(1-d_6 R^{-1/3} )| $. The scaling of the slowly rotating convection bound matches the zero rotation bound of \cite{Arslan2024}, noting that the initial assumptions on $\delta$ and $\varepsilon$ can be adjusted such that the constants in the two bounds match. The question of a bound dependent on $E$ that, in the limit of $E\rightarrow\infty$, matches the scaling of $R^{-2/3}$ remains open. In region II, $E\lesssim R^{-2}$ and $E_m\geq E \geq E_0$, the best bound would match I, but the region is below $R_L$ when $E=\infty$, so no convection occurs, and $\mathcal{F}_B=\frac12$. In region III, $E\gtrsim R^{-2}$ and $E\leq E_0$, the only bound is $ d_1 R^{-2/3}E^{2/3} + d_2 R^{-1/2}E^{1/2}|\ln(1-d_3 R^{-1/3}E^{1/3} )| $, provided $R>R_L$. Finally, in region IV, where
$E\lesssim R^{-2}$ and $E\leq E_0$, the best bound is $d_7 R^{-1} + d_8 R^{-4/5}|\ln(1-d_9 R^{-2/5})| $ again provided that  $R > R_L$. While region IV is below the dotted green line in \cref{fig:bounds_regimes}, \eqref{eq:R_E_rel} is an asymptotic result for instability with stress-free boundaries. The bounds, on the other hand, are for no-slip boundaries.

Moving on to \cref{fig:bounds_regimes}\textit{(b)} and the bounds \eqref{eq:thm3} and \eqref{eq:thm4}. The first difference with \textit{(a)} is that region I, is split into two.
In region I where $E\gtrsim R^{-2/3}$ and $E_m \geq E \geq E_1$, there are two bounds of $d_{13} R^{-1}E^{-1}$ and $d_{11} R^{-2/7}$, where from a comparison of the two the better bound is $d_{11} R^{-2/7}$, due to the requirement of $E\gtrsim R^{-2/3}$. However, for Ia where $E \gtrsim R^{-2/3}$ and $E\geq E_m$, the only valid bound is $d_{13} R^{-1}E^{-1}$. Region II, where $E\lesssim R^{-2/3}$ and $E \geq E_1$, is below the linear stability limit, so $\mean{T}=\tfrac{1}{12}$. In region III,  $E\gtrsim R^{-2/3}$ and $E\leq E_1$, there are two bounds of $d_{10} R^{-2/7}E^{2/7} $ and $d_{13} R^{-1}E^{-1}$, and both are valid in different cases. When $ R^{-2/3} \lesssim E\lesssim R^{-5/9}$, referred to as IIIa and shown with the purple region, the better lower bound is $d_{13}R^{-1}E^{-1}$, provided $R>R_L$. If instead $E\gtrsim R^{-5/9}$ then instead the mean temperature is bounded by $d_{10} R^{-2/7}E^{2/7} $.  The dashed green line scales as $R^{-3/4}$, so, for sufficiently large $R$, the entire region of III corresponds to convecting flows.
In region IV, $E\lesssim R^{-2/3}$ and $E\leq E_1$, there are two bounds but the better one is $d_{12} R^{-1/3}$, provided $R>R_L$. Table \ref{tab:bound_summary} summarises the above discussion of the results.
\begin{figure}
    \centering
    \includegraphics[scale=1]{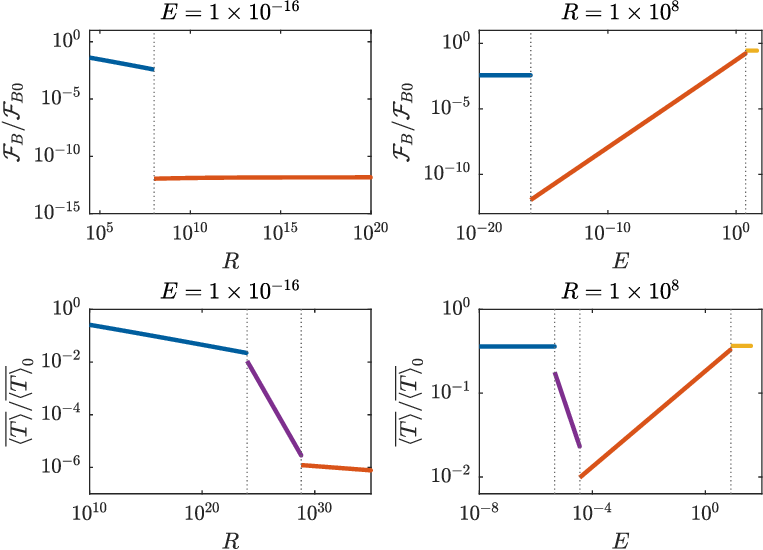}
     \begin{tikzpicture}[overlay]
        \node at (-11.2,8.6) {\textit{(a)}};
        \node at (-4.6,8.6) {\textit{(b)}};
        \node at (-11.2,3.85) {\textit{(c)}};
        \node at (-4.6,3.85) {\textit{(d)}};
    \end{tikzpicture}
    \caption{Plots of the ratio of lower bounds on $\mathcal{F}_B$ and $\mean{T}$ relative to the best known bounds at zero rotation $\mathcal{F}_{B0} $ and $\mean{T}_0$ \citep{Arslan2024,Whitehead2011}, for $E = 1\times 10^{-16}$ in \textit{(a)} and \textit{(c)} and $R=1\times 10^8$ in \textit{(b)} and \textit{(d)}. In all of the subplots, the blue lines ({\color{matlabblue}\solidrule}) are the lower bounds in region IV, the red lines ({\color{matlabred}\solidrule}) the bounds in region III and the yellow line ({\color{colorbar2}\solidrule}) the bounds in Region I, while purple lines in \textit{(c)} and \textit{(d)} are the bounds valid in regions Ia and IIIa for $\mean{T}$. The bounds themselves are in \cref{tab:bound_summary}. Vertical dotted lines show the intersections of the regions for each bound. }
    \label{fig:analysis1}
\end{figure}

Having established the regions where each of the bounds is valid in the $E-R$ space, we now consider the implications of these bounds on rotation in turbulent convection driven by internal heating. Figure \ref{fig:analysis1} compares the bounds proven in this paper with the best-known bounds for IHC without rotation \citep{Arslan2024, Whitehead2011}. We fix $E$ at $1 \times 10^{-16}$ and plot the bounds as a function of $R$, and then fix $R$ at $1 \times 10^{8}$ and plot them as a function of $E$.  Equivalent to taking horizontal and vertical slices of \cref{fig:bounds_regimes} to visualise the bounds along each line segment.

Discontinuities at the intersections in \cref{fig:analysis1}, defined by the regions of validity for each bound (see \cref{tab:bound_summary}), are expected and arise from comparing bounds obtained by different lemmas and are a product of the method used to ensure positivity of the spectral constraint of \eqref{e:spectral_constraint} in \S\ref{sec:Bounds_wT} and \S\ref{sec:Bounds_T}. Also, across all subplots, the ratio of the bounds is consistently less than 1, though by restricting the choices of $\delta$, $\varepsilon$, and estimates in the proofs, the $y$-values in \cref{fig:analysis1} can be made larger. For brevity in the proofs, constants were not optimised. Starting with $\mathcal{F}_B$ at $E = 10^{-16}$, the bounds in region III (red line) have the same scaling as the zero-rotation case, while in region IV (blue line), the bounds decrease, suggesting that at higher $R$, less heat may escape the domain compared to the non-rotating case.
For fixed $R$, only the bound in region III changes with $E$: for smaller $E$, the bound on $\mathcal{F}_B$ is smaller than $\mathcal{F}_{B0}$.
For $\mean{T}$, with $E$ fixed, the bounds scale such that the ratio $\mean{T}/\mean{T}_0$ always decreases. However, when $R$ is constant, the scaling in regions IV and I (yellow line) is independent of $E$. As rotation increases, the ratio decreases in region III but increases in regions Ia and IIIa (purple lines). A smaller lower bound in the rotating case implies a larger possible range for $\mathcal{F}_B$ and $\mean{T}$. Since rotation introduces new flow regimes, it is possible that some flows cause both quantities to be lower than in the absence of rotation.

Finally, consider the possible heuristic scalings of $\mean{T} \sim R^{-3/5} E^{-4/5}$ from  \eqref{eq:T_scaling2} and $\mathcal{F}_B \sim R^{-3/10} E^{-2/5} $ from \eqref{eq:wT_scaling2}, that may hold for rotation dominated convection. In the scaling laws, if we fix $R$ and take $E\rightarrow 0$, we tend to the uniform upper bounds for both quantities. The lower bounds closest to the heuristic scaling are those in region III (\cref{tab:bound_summary}). It is insightful to consider the small $E$ limit where $E \sim R_L^{-3/4}$, and we can write that
\begin{subequations}
\label{eq:heuristic_modify}
\begin{gather}
    \mean{T} \sim \left(\frac{R}{R_L} \right)^{-3/5}, \qquad \text{and} \\
    \mathcal{F}_B \sim \left(\frac{R}{R_L} \right)^{-3/10},
\end{gather}    
\end{subequations}
whereas the rigorous bounds in region III (\cref{tab:bound_summary}) become 
\begin{subequations}
\label{eq:bounds_modify}
\begin{gather}
    \mean{T} \gtrsim \left(\frac{R}{R_L} \right)^{3/14}R^{-1/2} , \qquad \text{and} \\
    \mathcal{F}_B \gtrsim \left(\frac{R}{R_L} \right)^{1/2} R^{-7/6} +  \left(\frac{R}{R_L} \right)^{3/8} R^{-7/8}\left| \ln{\left(1-\left(\frac{R}{R_L}\right)^{1/4} R^{-7/12} \right)} \right| .
\end{gather}    
\end{subequations}
Since $R$ is always a multiple of $R_L$, for $R>R_L$, the bounds in \eqref{eq:bounds_modify}, as expected, are smaller than the heuristic scaling laws \eqref{eq:heuristic_modify}. If the lower bounds are not sharp, then this would motivate the question of how to improve the bounds, which we now discuss with concluding remarks.

\subsection{Conclusions}

In this work we prove the first lower bounds on the mean temperature $\mean{T}$, (\eqref{eq:thm3} and \eqref{eq:thm4}) and the mean heat flux out of the bottom boundary $\mathcal{F}_B$ (\eqref{eq:thm_1} and \eqref{eq:thm2}), for rotating uniform internally heated convection (IHC) in the limit of infinite Prandtl number. Using the fact that the momentum equation in Rayleigh-B\'enard (RBC) and IHC is identical, we adapt estimates from \cite{Yan2004} and \cite{constantin1999r} to prove the first Rayleigh and Ekman number dependent bounds on $\mean{T}$ and $\mathcal{F}_B$ in IHC. By application of the auxiliary functional method, we prove bounds that apply to different regions of buoyancy to rotation-dominated flows, summarised in \cref{fig:bounds_regimes} and \cref{tab:bound_summary}. In addition to rigorous bounds, we demonstrate that the critical Rayleigh number for linear stability, $R_L$, asymptotically scales with the Ekman number as $E^{-4/3}$ when the marginally stable states are steady.

In contrast to previous applications of the background method, there are several unique features in the proofs of bounds in this work. Firstly, the background temperature fields have boundary layers of different widths for the proofs on $\mathcal{F}_B$, but not for $\mean{T}$. In particular, when we use \cref{lem:yan} we find that $\delta = \varepsilon^{2}$, while when using \cref{lem:const}, $\delta = \varepsilon^{5/2}$. 
The relation from using \cref{lem:yan} (\cref{sec:large_E_wt}) matches the predictions of heuristic arguments (\cref{sec:heuristics}). However, whether or not the background profiles are optimal remains unknown and can be addressed with numerical optimisation over a finite range of parameters \citep{Fantuzzi2016PRE,Fantuzzi2018,Fantuzzi2022}.
The scaling of the bound obtained for $\mathcal{F}_B$ in the slowly rotating region (region I in \cref{fig:bounds_regimes} and \cref{tab:bound_summary}) matches the best known bound for zero rotation, however, the constants differ. The scaling of the bounds on $\mean{T}$ do not match the zero rotation case, but the bounds in this work only hold up to $E\leq E_m = 41.4487$, and the background profile is not logarithmic, which is critical to the bound of \cite{Whitehead2011}. It would be interesting if further investigations can prove a bound that holds for large $E$ too and matches the scaling of the zero rotation bounds on $\mean{T}$ \citep{Whitehead2011} and $\mathcal{F}_B$ \citep{Arslan2024} when $E\rightarrow\infty$. The two lemmas used in this work give different bounds for different regions of the $E-R$ parameter. While both estimate the second derivative of the vertical velocity, the estimates hold for different $L^p$ norms with \cref{lem:yan} utilising a finer spectral analysis of the Greens function of \eqref{eq:mom_eq}. More generally, \cref{lem:yan} is a pointwise estimate in $z$, and \cref{lem:const} is an integral estimate over the whole domain.

Although rigorous demonstrations of the validity of the results at arbitrary $Pr$ are not provided here, previous work in \cite{tilgner2022} outlines a strategy for extending bounds from $Pr=\infty$ to finite $Pr$ for RBC. The author achieves this under specific restrictions on $E$ and a numerical approximation of a Greens' function, making the bound semi-analytic. 
A similar approach appears in \cite{wang2013} to extend bounds on RBC for stress-free boundaries from infinite $Pr$ to arbitrary $Pr$ in 3D (the proof in 2D being given in \cite{Whitehead2011prl}). The barrier to adapting to IHC is the lack of a maximum principle on the temperature of the form $\lVert T \rVert_\infty \leq c$, where $c=1$ for RBC. Proof of any maximum principle for IHC is unknown for any $L^p$ space. Therefore, at best, we can conjecture that akin to RBC, to highest order, the bounds in this work should hold for arbitrary $Pr$.

In considering turbulent convection subject to rotation, a question of interest is the behaviour in the limit of rapidly rotating convection. Rapidly rotating IHC could be investigated by taking the approach of the non-hydrostatic quasigeostrophic approximation \citep{julien1996,sprague2006,julien2016} to \eqref{eq:gov_eqs}. It is worth noting that the bounds for the rapidly rotating limit in RBC apply to arbitrary $Pr$, making the results relevant for geophysical flows. However, for geophysical applications, in addition to rapid rotation, IHC in spherical geometry is of importance. No rigorous study on the turbulent state of such a system is known, and any change in the bounds with a variation in the geometry is an intriguing avenue for future research.

For any result obtained with a bounding method, an important question is on the sharpness of the bounds. High-resolution numerical simulations can provide insight into the sharpness of the bounds in \eqref{tab:bound_summary}. A numerical study of the parameter space would provide valuable insight into the nature of heat transport in uniform rotating IHC, as no such data, numerical and experimental, exists to our knowledge on IHC. Then, proof of better bounds, by moving away from quadratic auxiliary functionals, and hence the background field method, can also answer the question of sharpness. In general, mathematical improvements are obtained in one of two ways: either by changes to the variational problem, such that the expressions for the bounds and the spectral constraint change, or by novel estimates of the flow quantities. The latter method is more mathematically challenging, while the prior can be achieved with new physical insights. For example, given that additional constraints, like minimum and maximum principles, improve bounds for convection \citep{Otto2011,Arslan2021}, it would be interesting to see if information about rotation can be exploited to construct a variational problem from \eqref{eq:gov_eqs}, that yields better bounds. Beyond the relation between the $R_L$ and $E$, a trait of rotating flows is the Taylor-Proudman theorem, which could form the basis of an additional constraint which improves the bounds. Also, alternative auxiliary functionals might bear fruit in studying bounds on rotating convection. More concretely, the linear stability analysis reveals the importance of the vertical vorticity, and functionals that incorporate vorticity may provide insight into improved bounds, especially given the conjecture in \cite{CHERNYSHENKO2023} for the use of helicity in the auxiliary functional. Similar functionals appear in studies on nonlinear stability of rotating RBC \citep{galdi1985,giacobbe2014}.

As a final remark, recent work has highlighted novel results when the internal heating is non-uniform \citep{Lepot2018,bouillaut2022,Song2022,Arslan2024b}. The change in the physics or bounds due to distributed heating or cooling would be an interesting future line of research. From the perspective of the PDE, extending the setup in \eqref{eq:nondim_energy} to arbitrary heating profiles would be the natural next step when studying bounds on the long-time behaviour of turbulent convection.

\backsection[Acknowledgments]{ The author thanks Andrew Jackson for valuable discussions, and insights and for commenting on the manuscript and Fabian Burnman for reading and commenting on the manuscript. }{}

\backsection[Funding]{The author acknowledges funding from the European Research Council (agreement no. 833848-UEMHP) under the Horizon 2020 program and the Swiss National Science Foundation (grant number 219247) under the MINT 2023 call.  }

\backsection[Declaration of interests]{ The author reports no conflict of interest.}

\backsection[Author ORCIDs]{ \\
https://orcid.org/0000-0002-5824-5604\\}

\appendix

\section{Table of constants}
\label{app:constant}

For clarity in the proofs of the lower bounds on $\mathcal{F}_B$ and $\mean{T}$ (\eqref{eq:thm_1}, \eqref{eq:thm2}, \eqref{eq:thm3} and \eqref{eq:thm4}), collated here are the constants that appear in the bounds. References to the precise equations where they appear are included.

    \begin{center}
    \begin{tabular}{cccc}
        \hline
         Bound&Eq. no.&Constant&Value \\\hline
         $\mathcal{F}_B$& \eqref{eq:thm_1} &$d_1$& $ \frac{18-7\sqrt{3}}{36} \left(\frac{7\sqrt{6}}{16 c_2}\right)^{2/3} $ \\
         &\eqref{eq:thm_1}&$d_2$& $\frac{1}{12} \left(\frac{1}{6}\right)^{1/12} \sqrt{\frac{21}{c_2}}  $ \\
         &\eqref{eq:thm_1}&$d_3$& $ \left(\frac{7\sqrt{6}}{12c_2}\right)^{1/3} $ \\
         &\eqref{eq:thm_1}&$d_4$& $\frac{18-7\sqrt{3}}{36} \left( \frac{49\sqrt{2}}{2c_2^2} \right)^{1/3}  $ \\
         &\eqref{eq:thm_1}&$d_5$& $\frac{\sqrt{7}}{3} \Big( \frac{9}{32\sqrt{2} c_2^5 } \Big)^{1/12} $ \\
         &\eqref{eq:thm_1}&$d_6$&  $ \left( \frac{392\sqrt{2}}{9c_2^2} \right)^{1/6} $\\
         &\eqref{eq:thm2}&$d_7$& $ \frac{\sqrt{3}\pi(18-7\sqrt{3})}{96}$ \\
         &\eqref{eq:thm2}&$d_8$& $ \frac{(\sqrt{3}\pi)^{4/5}}{2^{14/5}} $\\
         &\eqref{eq:thm2}&$d_9$& $ \left(\frac{\sqrt{3}\pi}{2}\right)^{2/5} $\\
         \hline
         $\mean{T}$& \eqref{eq:thm3} & $d_{10}$& $\frac29\left( \frac{7\sqrt{2}}{8c_2}\right)^{2/7}$\\
         &\eqref{eq:thm3}&$d_{11}$ & $  \frac29\left( \frac{196}{c_2^2} \right)^{1/7} $ \\
         &\eqref{eq:thm4}&$d_{12}$ & $ \frac29  \left( \frac{9\pi}{8} \right)^{1/3} $ \\
         &\eqref{eq:thm4}&$d_{13}$ &$ \frac16$ \\\hline
    \end{tabular}
    
    \label{tab:constants}
    \end{center}

\section{Heuristics scaling arguments}
\label{sec:heuristics}

Owing to the lack of data on uniform internally heated convection subject to rotation, we cannot comment on the sharpness of the bounds proven. Instead, we can use standard physical arguments to determine possible scaling laws for $\mean{T}$ and $\mathcal{F}_B$. In previous studies, the theory of Grossman \& Lohse \citep{GroSjfm2000a,Ahlers2009} has been adapted to uniform and exponentially varying IHC to determine scaling laws in the non-rotating case \citep{Wang2020,creyssels2020,creyssels_2021}. Here, we will follow the heuristic arguments presented in \cite{Arslan2021} that adapt the ideas of marginal stability and diffusivity-free scaling of \cite{Malkus1954} and \cite{spiegel1963generalization}. Heuristic arguments can be adapted to the rotating case to propose possible scaling laws for rotation-dominated convection 
\citep{king2009,aurnou2020,ecke2023}.

The starting point is to suppose that the heat fluxes out of the domain, defined in \eqref{eq:Flux_def}, can be written as $\mathcal{F}_B \sim \mean{T}/\delta$ and $\mathcal{F}_T \sim \mean{T}/\varepsilon$. Once again, $\delta$ and $\varepsilon$ are the thermal boundary layer thicknesses at the bottom and top and are different sizes. 
In this section, $\sim$ means approximately equal to up to constants.
Then, in the buoyancy and rotation-dominated regimes, we assume that to the highest order, the mean temperature is a function of the Rayleigh and Ekman numbers, more precisely
\begin{equation}
   \mean{T} \sim R^{-\alpha}, \quad \text{and} \quad \mean{T} \sim (R/R_L)^{-\gamma} = R^{-\gamma} E^{-4\gamma/3} \, ,
   \label{eq:heuristic_T}
\end{equation} 
where $\alpha \in \mathbb{R}_+$ and $\gamma \in \mathbb{R}_+$ are exponents to be determined and we have substituted for $R_L$ with \eqref{eq:R_E_rel}. 
The assumption \eqref{eq:heuristic_T} is justified in the non-rotating case by numerical studies of $\mean{T}$ \citep[Table 3.2 and references therein]{Goluskin2016book}.

The two main regimes of buoyancy and rotation dominated turbulent convection can be interpolated by varying the Rayleigh and Ekman numbers. However, in going from buoyancy to rotation-dominated heat transport, the thermal boundary layer will become larger than the Ekman boundary layer. However, first, we need to determine the behaviour of the thermal boundary layers. One possible argument, but by no means the only one, is the following. In the bottom boundary, heating balances diffusion, given that the flow is stably stratified. Then, heating over $\delta$ is proportional to $\delta$ while diffusion scales as $\mean{T}/\delta$, implying that $\delta^2 \sim \mean{T}$ and by the energy balance of $\mathcal{F}_T + \mathcal{F}_B = 1$ that $\delta^2 \sim \varepsilon$. Stated in words, the upper thermal boundary layer scales as the mean temperature and is the square of the lower thermal boundary layer. The implication is that $\mathcal{F}_B \sim \mean{T}^{1/2}$. Turning to the Ekman boundary layers, by standard arguments $\delta_E \sim E^{1/2}$ \citep{stevenson1979}. Therefore, using \eqref{eq:heuristic_T} and supposing $\delta \sim \delta_E$, the resulting algebraic equation gives
\begin{equation}
    \label{eq:gam_alf_rel}
    \gamma = \frac{3 \alpha}{3 - 4\alpha} .
\end{equation}
The relationship in \eqref{eq:gam_alf_rel} gives a range of possible scaling behaviours for the IHC as the flow transitions from buoyancy to rotation-dominated, and it then remains to determine $\alpha$. If we first rearrange \eqref{eq:gam_alf_rel} in terms of $\alpha$, we find that $\alpha = 3\gamma / (3+4\gamma)$. For $\gamma\rightarrow\infty$, then $\alpha\rightarrow\tfrac34$, and the maximal exponent of $\alpha$ is $\tfrac34$, but this does not correspond to any physical arguments and is ruled out by rigorous bounds \citep{Lu2004,Whitehead2011}.

It remains to determine $\alpha$ to obtain the desired heuristic scaling laws. If we use the argument of marginal stability \citep{Malkus1954,Howard1963} to the unstably stratified upper thermal boundary layer, $\varepsilon$, we find that $\alpha=\tfrac14$ and call this the classical exponent. If, instead, turbulent heat transport is independent of the fluid diffusivities and is given by a characteristic free-fall velocity \citep{spiegel1963generalization}, we find $\alpha = \tfrac13$ and refer to this as the ultimate exponent. See \cite{Arslan2021} for a detailed explanation of the exponents for IHC in the non-rotating case. Then, for $\alpha=\tfrac14$ or $\tfrac13$, using \eqref{eq:gam_alf_rel} gives the following predictions in the buoyancy and rotation-dominated regimes,
\begin{subequations}
\begin{align}
    \label{eq:T_scaling}
    \text{non-rotating}: \qquad\mean{T}& \sim
    \begin{dcases}
        R^{-1/4} , & ~~\text{classical}, \\
       R^{-1/3} , & ~~\text{ultimate},
    \end{dcases}\\
    \label{eq:T_scaling2}
    \text{rotating}: \qquad\mean{T}& \sim
    \begin{dcases}
        R^{-3/8}E^{-1/2} , & ~~\text{classical}, \\
       R^{-3/5} E^{-4/5}, & ~~\text{ultimate},
    \end{dcases}
\end{align}
\end{subequations}
and
\begin{subequations}
\begin{align}
    \label{eq:wT_scaling}
    \text{non-rotating}: \qquad\mathcal{F}_B& \sim
    \begin{dcases}
        R^{-1/8} , & ~~\text{classical}, \\
       R^{-1/6} , & ~~\text{ultimate},
    \end{dcases}\\
    \label{eq:wT_scaling2}
    \text{rotating}: \qquad\mathcal{F}_B& \sim
    \begin{dcases}
        R^{-3/16}E^{-1/4} , & ~~\text{classical}, \\
       R^{-3/10} E^{-2/5}, & ~~\text{ultimate}.
    \end{dcases}
\end{align}
\end{subequations}
While the classical regime for a rotating flow is not physically relevant \citep{ecke2023}, it appears in \eqref{eq:T_scaling2} and \eqref{eq:wT_scaling2} for completeness.

As mentioned in the introduction, one can define a proxy Nusselt number as $Nu_p = 1/\mean{T}$.
Furthermore, the temperature difference based Rayleigh number, $Ra$, appearing in studies of boundary driven thermal convection, is related to the flux-based Rayleigh number, $R$, through the relation that $Nu_p = R/Ra$. Therefore, substituting for $R$ in the scaling laws \eqref{eq:T_scaling} and \eqref{eq:T_scaling2} returns the known scaling laws for the Nusselt number in RBC of $Nu\sim Ra^{1/2}$ and $Nu \sim $ $ Ra^{3/2}E^2$ for the ultimate scaling. 
We comment on the heuristic scaling laws in \cref{sec:conclusion} and compare them to the bounds we prove in the subsequent sections.

\bibliographystyle{jfm}
\bibliography{jfm}

\begin{thebibliography}{97}
\expandafter\ifx\csname natexlab\endcsname\relax\def\natexlab#1{#1}\fi
\def\au#1{#1} \def\ed#1{#1} \def\yr#1{#1}\def\at#1{#1}\def\jt#1{\textit{#1}} \def\bt#1{#1}\def\bvol#1{\textbf{#1}} \def\vol#1{#1} \def\pg#1{#1} \def\publ#1{#1}\def\arxiv#1{#1}\def\org#1{#1}\def\st#1{\textit{#1}}

\bibitem[Ahlers {\em et~al.\/}(2009)Ahlers, Grossmann \& Lohse]{Ahlers2009}
{\sc \au{Ahlers, G.}, \au{Grossmann, S.} \& \au{Lohse, D.}} \yr{2009}  \at{{Heat transfer and large scale dynamics in turbulent Rayleigh--B\'enard convection}}.  \jt{Reviews of Modern Physics}  \bvol{81}~(2),  \pg{503--537}.

\bibitem[Arslan {\em et~al.\/}(2021{\natexlab{{\em a\/}}})Arslan, Fantuzzi, Craske \& Wynn]{Arslan2021a}
{\sc \au{Arslan, A.}, \au{Fantuzzi, G.}, \au{Craske, J.} \& \au{Wynn, A.}} \yr{2021{\natexlab{{\em a\/}}}}  \at{{Bounds for internally heated convection with fixed boundary heat flux}}.  \jt{Journal of Fluid Mechanics}  \bvol{992},  \pg{R1}.

\bibitem[Arslan {\em et~al.\/}(2021{\natexlab{{\em b\/}}})Arslan, Fantuzzi, Craske \& Wynn]{Arslan2021}
{\sc \au{Arslan, A.}, \au{Fantuzzi, G.}, \au{Craske, J.} \& \au{Wynn, A.}} \yr{2021{\natexlab{{\em b\/}}}}  \at{{Bounds on heat transport for convection driven by internal heating}}.  \jt{Journal of Fluid Mechanics}  \bvol{919},  \pg{A15}.

\bibitem[Arslan {\em et~al.\/}(2023)Arslan, Fantuzzi, Craske \& Wynn]{Arslan2023}
{\sc \au{Arslan, A.}, \au{Fantuzzi, G.}, \au{Craske, J.} \& \au{Wynn, A.}} \yr{2023}  \at{Rigorous scaling laws for internally heated convection at infinite prandtl number}.  \jt{Journal of Mathematical Physics}  \bvol{64}~(2),  \pg{023101}.

\bibitem[Arslan {\em et~al.\/}(2024)Arslan, Fantuzzi, Craske \& Wynn]{Arslan2024b}
{\sc \au{Arslan, A.}, \au{Fantuzzi, G.}, \au{Craske, J.} \& \au{Wynn, A.}} \yr{2024}  \at{Internal heating profiles for which downward conduction is impossible}.  \jt{Journal of Fluid Mechanics}  \bvol{993},  \pg{A5}.

\bibitem[Arslan \& Rojas(2024)]{Arslan2024}
{\sc \au{Arslan, A.} \& \au{Rojas, R.~E.}} \yr{2024}  \at{{New bounds for heat transport in internally heated convection at infinite Prandtl number}}.  \jt{arXiv:2403.14407 [physics.flu-dyn]} .

\bibitem[Aurnou {\em et~al.\/}(2020)Aurnou, Horn \& Julien]{aurnou2020}
{\sc \au{Aurnou, J.~M.}, \au{Horn, S.} \& \au{Julien, K.}} \yr{2020}  \at{Connections between nonrotating, slowly rotating, and rapidly rotating turbulent convection transport scalings}.  \jt{Physical Review Research}  \bvol{2}~(4),  \pg{043115}.

\bibitem[Barker {\em et~al.\/}(2014)Barker, Dempsey \& Lithwick]{barker2014}
{\sc \au{Barker, A.~J.}, \au{Dempsey, A.~M.} \& \au{Lithwick, Y.}} \yr{2014}  \at{Theory and simulations of rotating convection}.  \jt{The Astrophysical Journal}  \bvol{791}~(1),  \pg{13}.

\bibitem[Boubnov \& Golitsyn(2012)]{boubnov2012}
{\sc \au{Boubnov, B.~M.} \& \au{Golitsyn, G.~S.}} \yr{2012} {\em Convection in rotating fluids\/}, ,  \vol{vol.~29}.  \publ{Springer Science \& Business Media}.

\bibitem[Bouillaut {\em et~al.\/}(2022)Bouillaut, Flesselles, Miquel, Aumaître \& Gallet]{bouillaut2022}
{\sc \au{Bouillaut, V.}, \au{Flesselles, B.}, \au{Miquel, B.}, \au{Aumaître, S.} \& \au{Gallet, B.}} \yr{2022}  \at{Velocity-informed upper bounds on the convective heat transport induced by internal heat sources and sinks}.  \jt{Philosophical Transactions of the Royal Society A}  \bvol{380}~(2225),  \pg{20210034}.

\bibitem[Busse(1970)]{Busse1970}
{\sc \au{Busse, F.~H.}} \yr{1970}  \at{{Bounds for turbulent shear flow}}.  \jt{Journal of Fluid Mechanics}  \bvol{41}~(1),  \pg{219--240}.

\bibitem[Chandrasekhar(1961)]{chandrasekhar2013hydrodynamic}
{\sc \au{Chandrasekhar, S.}} \yr{1961} {\em Hydrodynamic and hydromagnetic stability\/}.  \publ{Oxford University Press}.

\bibitem[Chernyshenko(2022)]{Chernyshenko2022}
{\sc \au{Chernyshenko, S.}} \yr{2022}  \at{{Relationship between the methods of bounding time averages}}.  \jt{Philosophical Transactions of the Royal Society A}  \bvol{380}~(1),  \pg{20210044.}

\bibitem[Chernyshenko(2023)]{CHERNYSHENKO2023}
{\sc \au{Chernyshenko, S.}} \yr{2023}  \at{Background flow hidden in a bound for nusselt number}.  \jt{Physica D: Nonlinear Phenomena}  \bvol{445},  \pg{133641}.

\bibitem[Chernyshenko {\em et~al.\/}(2014)Chernyshenko, Goulart, Huang \& Papachristodoulou]{Chernyshenko2014a}
{\sc \au{Chernyshenko, S.}, \au{Goulart, P.~J.}, \au{Huang, D.} \& \au{Papachristodoulou, A.}} \yr{2014}  \at{{Polynomial sum of squares in fluid dynamics: a review with a look ahead}}.  \jt{Philosophical Transactions of the Royal Society A}  \bvol{372}~(2020),  \pg{20130350}.

\bibitem[Constantin(1994)]{constantin1994}
{\sc \au{Constantin, P.}} \yr{1994}  \at{{Geometric statistics in turbulence}}.  \jt{SIAM Review.}  \bvol{36}~(1),  \pg{73--98}.

\bibitem[Constantin \& Doering(1995)]{Constantin1995a}
{\sc \au{Constantin, P.} \& \au{Doering, C.~R.}} \yr{1995}  \at{{Variational bounds on energy dissipation in incompressible flows. II. Channel flow}}.  \jt{Physical Review E}  \bvol{51}~(4),  \pg{3192--3198}.

\bibitem[Constantin {\em et~al.\/}(2001)Constantin, Hallstrom \& Poutkaradze]{constantin2001}
{\sc \au{Constantin, P.}, \au{Hallstrom, C.} \& \au{Poutkaradze, V.}} \yr{2001}  \at{Logarithmic bounds for infinite prandtl number rotating convection}.  \jt{Journal of Mathematical Physics}  \bvol{42}~(2),  \pg{773--783}.

\bibitem[Constantin {\em et~al.\/}(1999)Constantin, Hallstrom \& Putkaradze]{constantin1999r}
{\sc \au{Constantin, P.}, \au{Hallstrom, C.} \& \au{Putkaradze, V.}} \yr{1999}  \at{Heat transport in rotating convection}.  \jt{Physica D: Nonlinear Phenomena}  \bvol{125}~(3-4),  \pg{275--284}.

\bibitem[Creyssels(2020)]{creyssels2020}
{\sc \au{Creyssels, M.}} \yr{2020}  \at{Model for classical and ultimate regimes of radiatively driven turbulent convection}.  \jt{Journal of Fluid Mechanics}  \bvol{900},  \pg{A39}.

\bibitem[Creyssels(2021)]{creyssels_2021}
{\sc \au{Creyssels, M.}} \yr{2021}  \at{Model for thermal convection with uniform volumetric energy sources}.  \jt{Journal of Fluid Mechanics}  \bvol{919},  \pg{A13}.

\bibitem[Currie {\em et~al.\/}(2020)Currie, Barker, Lithwick \& Browning]{currie2020}
{\sc \au{Currie, L.~K.}, \au{Barker, A.~J.}, \au{Lithwick, Y.} \& \au{Browning, M.~K}} \yr{2020}  \at{Convection with misaligned gravity and rotation: Simulations and rotating mixing length theory}.  \jt{Monthly Notices of the Royal Astronomical Society}  \bvol{493}~(4),  \pg{5233--5256}.

\bibitem[Davis(1969)]{Davis1969}
{\sc \au{Davis, S.~H.}} \yr{1969}  \at{On the principle of exchange of stabilities}.  \jt{Proceedings of the Royal Society of London. Series A, Mathematical and Physical Sciences}  \bvol{310}~(1502),  \pg{341--358}.

\bibitem[Ding \& Marensi(2019)]{ding2019tc}
{\sc \au{Ding, Z.} \& \au{Marensi, E.}} \yr{2019}  \at{Upper bound on angular momentum transport in taylor-couette flow}.  \jt{Physical Review E}  \bvol{100}~(6),  \pg{063109}.

\bibitem[Doering(2020)]{doering2020turning}
{\sc \au{Doering, C.~R.}} \yr{2020}  \at{Turning up the heat in turbulent thermal convection}.  \jt{Proceedings of the National Academy of Sciences}  \bvol{117}~(18),  \pg{9671--9673}.

\bibitem[Doering \& Constantin(1992)]{doering1992energy}
{\sc \au{Doering, C.~R.} \& \au{Constantin, P.}} \yr{1992}  \at{{Energy dissipation in shear driven turbulence}}.  \jt{Physical Review Letters}  \bvol{69}~(11),  \pg{1648}.

\bibitem[Doering \& Constantin(1994)]{Doering1994}
{\sc \au{Doering, C.~R.} \& \au{Constantin, P.}} \yr{1994}  \at{{Variational bounds on energy dissipation in incompressible flows: Shear flow}}.  \jt{Physical Review E}  \bvol{49}~(5),  \pg{4087--4099}.

\bibitem[Doering \& Constantin(1996)]{Doering1996}
{\sc \au{Doering, C.~R.} \& \au{Constantin, P.}} \yr{1996}  \at{{Variational bounds on energy dissipation in incompressible flows. III. Convection}}.  \jt{Physical Review E}  \bvol{53}~(6),  \pg{5957--5981}.

\bibitem[Doering \& Constantin(2001)]{Doering2001}
{\sc \au{Doering, C.~R.} \& \au{Constantin, P.}} \yr{2001}  \at{{On upper bounds for infinite Prandtl number convection with or without rotation}}.  \jt{Journal of Mathematical Physics}  \bvol{42}~(2),  \pg{784--795}.

\bibitem[Doering {\em et~al.\/}(2006)Doering, Otto \& Reznikoff]{Doering2006}
{\sc \au{Doering, C.~R.}, \au{Otto, F.} \& \au{Reznikoff, M.~G.}} \yr{2006}  \at{{Bounds on vertical heat transport for infinite Prandtl number Rayleigh--B\'enard convection}}.  \jt{Journal of Fluid Mechanics}  \bvol{560},  \pg{229--241}.

\bibitem[Ecke \& Shishkina(2023)]{ecke2023}
{\sc \au{Ecke, R.~E.} \& \au{Shishkina, O.}} \yr{2023}  \at{Turbulent rotating rayleigh--b{\'e}nard convection}.  \jt{Annual review of fluid mechanics}  \bvol{55},  \pg{603--638}.

\bibitem[Fantuzzi(2018)]{Fantuzzi2018}
{\sc \au{Fantuzzi, G.}} \yr{2018}  \at{{Bounds for Rayleigh--B{\'{e}}nard convection between free-slip boundaries with an imposed heat flux}}.  \jt{Journal of Fluid Mechanics}  \bvol{837},  \pg{R5}.

\bibitem[Fantuzzi {\em et~al.\/}(2022)Fantuzzi, Arslan \& Wynn]{Fantuzzi2022}
{\sc \au{Fantuzzi, G.}, \au{Arslan, A.} \& \au{Wynn, A.}} \yr{2022}  \at{{The background method: Theory and computations}}.  \jt{Philosophical Transactions of the Royal Society A}  \bvol{380}~(1),  \pg{20210038}.

\bibitem[Fantuzzi \& Wynn(2016)]{Fantuzzi2016PRE}
{\sc \au{Fantuzzi, G.} \& \au{Wynn, A.}} \yr{2016}  \at{{Optimal bounds with semidefinite programming: An application to stress-driven shear flows}}.  \jt{Physical Review E}  \bvol{93}~(4),  \pg{043308}.

\bibitem[Galdi \& Straughan(1985)]{galdi1985}
{\sc \au{Galdi, G.~P.} \& \au{Straughan, B.}} \yr{1985}  \at{A nonlinear analysis of the stabilizing effect of rotation in the b{\'e}nard problem}.  \jt{Proceedings of the Royal Society of London. A. Mathematical and Physical Sciences}  \bvol{402}~(1823),  \pg{257--283}.

\bibitem[Giacobbe \& Mulone(2014)]{giacobbe2014}
{\sc \au{Giacobbe, A.} \& \au{Mulone, G.}} \yr{2014}  \at{Stability in the rotating benard problem and its optimal lyapunov functions}.  \jt{Acta Applicandae Mathematicae}  \bvol{132},  \pg{307--320}.

\bibitem[Glatzmaier(2013)]{glatzmaier2013}
{\sc \au{Glatzmaier, G.}} \yr{2013} {\em Introduction to modeling convection in planets and stars: Magnetic field, density stratification, rotation\/}.  \publ{Princeton University Press}.

\bibitem[Goluskin(2015)]{Goluskin2016book}
{\sc \au{Goluskin, D.}} \yr{2015} {\em {Internally heated convection and Rayleigh--B\'enard convection}\/}.  \publ{Springer}.

\bibitem[Goluskin \& van~der Poel(2016)]{goluskin2016penetrative}
{\sc \au{Goluskin, D.} \& \au{van~der Poel, E.~P.}} \yr{2016}  \at{{Penetrative internally heated convection in two and three dimensions}}.  \jt{Journal of Fluid Mechanics}  \bvol{791}.

\bibitem[Goluskin \& Spiegel(2012)]{goluskin2012convection}
{\sc \au{Goluskin, D.} \& \au{Spiegel, E.~A.}} \yr{2012}  \at{{Convection driven by internal heating}}.  \jt{Physics Letters A}  \bvol{377}~(1-2),  \pg{83--92}.

\bibitem[Greenspan(1968)]{greenspan1968}
{\sc \au{Greenspan, H.~P.}} \yr{1968} {\em The theory of rotating fluids\/}.  \publ{Cambridge University Press}.

\bibitem[Grooms {\em et~al.\/}(2010)Grooms, Julien, Weiss \& Knobloch]{grooms2010}
{\sc \au{Grooms, I.}, \au{Julien, K.}, \au{Weiss, J.~B.} \& \au{Knobloch, E.}} \yr{2010}  \at{Model of convective taylor columns in rotating rayleigh-b{\'e}nard convection}.  \jt{Physical review letters}  \bvol{104}~(22),  \pg{224501}.

\bibitem[Grooms \& Whitehead(2014)]{grooms2014}
{\sc \au{Grooms, I.} \& \au{Whitehead, J.~P.}} \yr{2014}  \at{Bounds on heat transport in rapidly rotating rayleigh--b{\'e}nard convection}.  \jt{Nonlinearity}  \bvol{28}~(1),  \pg{29}.

\bibitem[Grossmann \& Lohse(2000)]{GroSjfm2000a}
{\sc \au{Grossmann, S.} \& \au{Lohse, D.}} \yr{2000}  \at{Scaling in thermal convection: a unifying theory}.  \jt{Journal of Fluid Mechanics}  \bvol{407},  \pg{27–56}.

\bibitem[Guervilly \& Cardin(2016)]{guervilly2016}
{\sc \au{Guervilly, C.} \& \au{Cardin, P.}} \yr{2016}  \at{Subcritical convection of liquid metals in a rotating sphere using a quasi-geostrophic model}.  \jt{Journal of Fluid Mechanics}  \bvol{808},  \pg{61--89}.

\bibitem[Guzm{\'a}n {\em et~al.\/}(2020)Guzm{\'a}n, Madonia, Cheng, Ostilla-M{\'o}nico, Clercx \& Kunnen]{guzman2020}
{\sc \au{Guzm{\'a}n, A. J.~A.}, \au{Madonia, M.}, \au{Cheng, J.~S.}, \au{Ostilla-M{\'o}nico, R.}, \au{Clercx, H. J.~H.} \& \au{Kunnen, R. P.~J.}} \yr{2020}  \at{Competition between ekman plumes and vortex condensates in rapidly rotating thermal convection}.  \jt{Physical Review Letters}  \bvol{125}~(21),  \pg{214501}.

\bibitem[Hadjerci {\em et~al.\/}(2024)Hadjerci, Bouillaut, Miquel \& Gallet]{hadjerci2024}
{\sc \au{Hadjerci, G.}, \au{Bouillaut, V.}, \au{Miquel, B.} \& \au{Gallet, B.}} \yr{2024}  \at{{Rapidly rotating radiatively driven convection: experimental and numerical validation of teh 'geostrophuc turbulence' scaling predictions}} ,  \arxiv{arXiv: 2401.16200}.

\bibitem[Herant {\em et~al.\/}(1994)Herant, Benz, Hix, Fryer \& Colgate]{herant1994}
{\sc \au{Herant, M.}, \au{Benz, W.}, \au{Hix, W.~R.}, \au{Fryer, C.~L.} \& \au{Colgate, S.~A.}} \yr{1994}  \at{Inside the supernova: A powerful convective engine}.  \jt{Astrophysical Journal, Part 1 (ISSN 0004-637X), vol. 435, no. 1, p. 339-361}  \bvol{435},  \pg{339--361}.

\bibitem[Herron(2003)]{herron2003}
{\sc \au{Herron, I.~H.}} \yr{2003}  \at{On the principle of exchange of stabilities in rayleigh-b{\'e}nard convection, ii-no-slip boundary conditions}.  \jt{Annali dell’Universit{\`a} di Ferrara}  \bvol{49},  \pg{169--182}.

\bibitem[Howard(1963)]{Howard1963}
{\sc \au{Howard, L.~N.}} \yr{1963}  \at{{Heat transport by turbulent convection}}.  \jt{Journal of Fluid Mechanics}  \bvol{17}~(3),  \pg{405--432}.

\bibitem[Jones \& Schubert(2015)]{jones2015}
{\sc \au{Jones, C.~A.} \& \au{Schubert, G.}} \yr{2015}  \at{Thermal and compositional convection in the outer core}.  \jt{Treatise in Geophysics, Core Dynamics}  \bvol{8},  \pg{131--185}.

\bibitem[Jones {\em et~al.\/}(2000)Jones, Soward \& Mussa]{jones2000}
{\sc \au{Jones, C.~A.}, \au{Soward, A.~M.} \& \au{Mussa, A.~I.}} \yr{2000}  \at{The onset of thermal convection in a rapidly rotating sphere}.  \jt{Journal of Fluid Mechanics}  \bvol{405},  \pg{157--179}.

\bibitem[Julien {\em et~al.\/}(2016)Julien, Aurnou, Calkins, Knobloch, Marti, Stellmach \& Vasil]{julien2016}
{\sc \au{Julien, K.}, \au{Aurnou, J.~M.}, \au{Calkins, M.~A.}, \au{Knobloch, E.}, \au{Marti, P.}, \au{Stellmach, S.} \& \au{Vasil, G.~M.}} \yr{2016}  \at{{A nonlinear model for rotationally constrained convection with Ekman pumping}}.  \jt{Journal of Fluid Mechanics}  \bvol{798},  \pg{50--87}.

\bibitem[Julien {\em et~al.\/}(1996)Julien, Legg, McWilliams \& Werne]{julien1996}
{\sc \au{Julien, K.}, \au{Legg, S.}, \au{McWilliams, J.} \& \au{Werne, J.}} \yr{1996}  \at{Rapidly rotating turbulent rayleigh-b{\'e}nard convection}.  \jt{Journal of Fluid Mechanics}  \bvol{322},  \pg{243--273}.

\bibitem[Julien {\em et~al.\/}(2012)Julien, Rubio, Grooms \& Knobloch]{julien2012}
{\sc \au{Julien, K.}, \au{Rubio, A.~M.}, \au{Grooms, I.} \& \au{Knobloch, E.}} \yr{2012}  \at{Statistical and physical balances in low rossby number rayleigh--b{\'e}nard convection}.  \jt{Geophysical \& Astrophysical Fluid Dynamics}  \bvol{106}~(4-5),  \pg{392--428}.

\bibitem[Kaplan {\em et~al.\/}(2017)Kaplan, Schaeffer, Vidal \& Cardin]{kaplan2017}
{\sc \au{Kaplan, E.~J.}, \au{Schaeffer, N.}, \au{Vidal, J.} \& \au{Cardin, P.}} \yr{2017}  \at{Subcritical thermal convection of liquid metals in a rapidly rotating sphere}.  \jt{Physical review letters}  \bvol{119}~(9),  \pg{094501}.

\bibitem[King {\em et~al.\/}(2013)King, Stellmach \& Buffett]{king2013}
{\sc \au{King, E.~M.}, \au{Stellmach, S.} \& \au{Buffett, B.}} \yr{2013}  \at{Scaling behaviour in rayleigh--b{\'e}nard convection with and without rotation}.  \jt{Journal of Fluid Mechanics}  \bvol{717},  \pg{449--471}.

\bibitem[King {\em et~al.\/}(2009)King, Stellmach, Noir, Hansen \& Aurnou]{king2009}
{\sc \au{King, E.~M.}, \au{Stellmach, S.}, \au{Noir, J.}, \au{Hansen, U.} \& \au{Aurnou, J.~M.}} \yr{2009}  \at{Boundary layer control of rotating convection systems}.  \jt{Nature}  \bvol{457}~(7227),  \pg{301--304}.

\bibitem[Knobloch(1998)]{knobloch1998}
{\sc \au{Knobloch, E.}} \yr{1998}  \at{Rotating convection: recent developments}.  \jt{International journal of engineering science}  \bvol{36}~(12-14),  \pg{1421--1450}.

\bibitem[Kumar(2022)]{Kumar_2022tc}
{\sc \au{Kumar, A.}} \yr{2022}  \at{Geometrical dependence of optimal bounds in taylor–couette flow}.  \jt{Journal of Fluid Mechanics}  \bvol{948},  \pg{A11}.

\bibitem[Kumar {\em et~al.\/}(2022)Kumar, Arslan, Fantuzzi, Craske \& Wynn]{kumar2021ihc}
{\sc \au{Kumar, A.}, \au{Arslan, A.}, \au{Fantuzzi, G.}, \au{Craske, J.} \& \au{Wynn, A.}} \yr{2022}  \at{Analytical bounds on the heat transport in internally heated convection}.  \jt{Journal of Fluid Mechanics}  \bvol{938},  \pg{A26}.

\bibitem[Kunnen(2021)]{kunnen2021}
{\sc \au{Kunnen, R. P.~J.}} \yr{2021}  \at{The geostrophic regime of rapidly rotating turbulent convection}.  \jt{Journal of Turbulence}  \bvol{22}~(4-5),  \pg{267--296}.

\bibitem[Lepot {\em et~al.\/}(2018)Lepot, Auma{\^{i}}tre \& Gallet]{Lepot2018}
{\sc \au{Lepot, S.}, \au{Auma{\^{i}}tre, S.} \& \au{Gallet, B.}} \yr{2018}  \at{{Radiative heating achieves the ultimate regime of thermal convection}}.  \jt{Proceedings of the National Academy of Sciences of the U.S.A.}  \bvol{115}~(36),  \pg{8937--8941}.

\bibitem[Lu {\em et~al.\/}(2004)Lu, Doering \& Busse]{Lu2004}
{\sc \au{Lu, L.}, \au{Doering, C.~R.} \& \au{Busse, F.~H.}} \yr{2004}  \at{{Bounds on convection driven by internal heating}}.  \jt{Journal of Mathematical Physics}  \bvol{45}~(7),  \pg{2967--2986}.

\bibitem[Malkus(1954)]{Malkus1954}
{\sc \au{Malkus, W. V.~R.}} \yr{1954}  \at{{The heat transport and spectrum of thermal turbulence}}.  \jt{Proceedings of the Royal Society A}  \bvol{225}~(1161),  \pg{196--212}.

\bibitem[Mulyukova \& Bercovici(2020)]{mulyukova2020}
{\sc \au{Mulyukova, E.} \& \au{Bercovici, D.}} \yr{2020}  \at{Mantle convection in terrestrial planets}.  \jt{Oxford Research Encyclopedia of Planetery Sciences} .

\bibitem[Nobili(2023)]{Nobili2022}
{\sc \au{Nobili, C.}} \yr{2023}  \at{The role of boundary conditions in scaling laws for turbulent heat transport}.  \jt{Mathematics in Engineering}  \bvol{5}~(1),  \pg{1--41}.

\bibitem[Otto \& Seis(2011)]{Otto2011}
{\sc \au{Otto, F.} \& \au{Seis, C.}} \yr{2011}  \at{{Rayleigh--B\'enard convection: Improved bounds on the Nusselt number}}.  \jt{Journal of Mathematical Physics}  \bvol{52}~(8),  \pg{083702}.

\bibitem[Pachev {\em et~al.\/}(2020)Pachev, Whitehead, Fantuzzi \& Grooms]{Pachev2020}
{\sc \au{Pachev, B.}, \au{Whitehead, J.~P.}, \au{Fantuzzi, G.} \& \au{Grooms, I.}} \yr{2020}  \at{{Rigorous bounds on the heat transport of rotating convection with Ekman pumping}}.  \jt{Journal of Mathematical Physics}  \bvol{61}~(2),  \pg{023101}.

\bibitem[Plumley \& Julien(2019)]{plumley2019}
{\sc \au{Plumley, M.} \& \au{Julien, K.}} \yr{2019}  \at{Scaling laws in rayleigh-b{\'e}nard convection}.  \jt{Earth and Space Science}  \bvol{6}~(9),  \pg{1580--1592}.

\bibitem[Radice {\em et~al.\/}(2016)Radice, Ott, Abdikamalov, Couch, Haas \& Schnetter]{radice2016}
{\sc \au{Radice, D.}, \au{Ott, C.~D.}, \au{Abdikamalov, E.}, \au{Couch, S.~M.}, \au{Haas, R.} \& \au{Schnetter, E.}} \yr{2016}  \at{Neutrino-driven convection in core-collapse supernovae: high-resolution simulations}.  \jt{The Astrophysical Journal}  \bvol{820}~(1),  \pg{76}.

\bibitem[Roberts(1967)]{roberts1967convection}
{\sc \au{Roberts, P.~H.}} \yr{1967}  \at{Convection in horizontal layers with internal heat generation. theory}.  \jt{Journal of Fluid Mechanics}  \bvol{30}~(1),  \pg{33--49}.

\bibitem[Roberts(1968)]{roberts1968thermal}
{\sc \au{Roberts, P.~H.}} \yr{1968}  \at{On the thermal instability of a rotating-fluid sphere containing heat sources}.  \jt{Philosophical Transactions of the Royal Society of London A}  \bvol{263}~(1136),  \pg{93--117}.

\bibitem[Rosa \& Temam(2022)]{Rosa2020}
{\sc \au{Rosa, R. M.~S.} \& \au{Temam, R.~M.}} \yr{2022}  \at{{Optimal minimax bounds for time and ensemble averages of dissipative infinite-dimensional systems with applications to the incompressible Navier--Stokes equations}}.  \jt{Pure and Applied Functional Analysis}  \bvol{7}~(1),  \pg{327--355}.

\bibitem[Rossby(1969)]{rossby1969}
{\sc \au{Rossby, H.~T.}} \yr{1969}  \at{A study of b{\'e}nard convection with and without rotation}.  \jt{Journal of Fluid Mechanics}  \bvol{36}~(2),  \pg{309--335}.

\bibitem[Schubert(2015)]{schubert2015treatise}
{\sc \au{Schubert, G.}} \yr{2015} {\em Treatise on geophysics\/}.  \publ{Elsevier}.

\bibitem[Schubert {\em et~al.\/}(2001)Schubert, Turcotte \& Olson]{schubert2001mantle}
{\sc \au{Schubert, G.}, \au{Turcotte, D.~L.} \& \au{Olson, P.}} \yr{2001} {\em Mantle convection in the Earth and planets\/}.  \publ{Cambridge University Press}.

\bibitem[Schumacher \& Sreenivasan(2020)]{schumacher2020}
{\sc \au{Schumacher, J.} \& \au{Sreenivasan, K.~R.}} \yr{2020}  \at{Colloquium: Unusual dynamics of convection in the sun}.  \jt{Rev. Mod. Phys.}  \bvol{92},  \pg{041001}.

\bibitem[Song {\em et~al.\/}(2022)Song, Fantuzzi \& Tobasco]{Song2022}
{\sc \au{Song, B.}, \au{Fantuzzi, G.} \& \au{Tobasco, I.}} \yr{2022}  \at{Bounds on heat transfer by incompressible flows between balanced sources and sinks}.  \jt{Physica D: Nonlinear Phenomena}  \pg{p. 133591}.

\bibitem[Song {\em et~al.\/}(2024)Song, Shishkina \& Zhu]{Song2024}
{\sc \au{Song, J.}, \au{Shishkina, O.} \& \au{Zhu, X.}} \yr{2024}  \at{Scaling regimes in rapidly rotating thermal convection at extreme rayleigh numbers}.  \jt{Journal of Fluid Mechanics}  \bvol{984},  \pg{A45}.

\bibitem[Spiegel(1963)]{spiegel1963generalization}
{\sc \au{Spiegel, E.~A.}} \yr{1963}  \at{A generalization of the mixing-length theory of turbulent convection.}  \jt{The Astrophysical Journal}  \bvol{138},  \pg{216}.

\bibitem[Sprague {\em et~al.\/}(2006)Sprague, Julien, Knobloch \& Werne]{sprague2006}
{\sc \au{Sprague, M.}, \au{Julien, K.}, \au{Knobloch, E.} \& \au{Werne, J.}} \yr{2006}  \at{Numerical simulation of an asymptotically reduced system for rotationally constrained convection}.  \jt{Journal of Fluid Mechanics}  \bvol{551},  \pg{141--174}.

\bibitem[Stellmach {\em et~al.\/}(2014)Stellmach, Lischper, Julien, Vasil, Cheng, Ribeiro, King \& Aurnou]{stellmach2014}
{\sc \au{Stellmach, S.}, \au{Lischper, M.}, \au{Julien, K.}, \au{Vasil, G.}, \au{Cheng, J.~S.}, \au{Ribeiro, A.}, \au{King, E.~M.} \& \au{Aurnou, J.~M.}} \yr{2014}  \at{Approaching the asymptotic regime of rapidly rotating convection: boundary layers versus interior dynamics}.  \jt{Physical review letters}  \bvol{113}~(25),  \pg{254501}.

\bibitem[Stevens {\em et~al.\/}(2013)Stevens, van~der Poel, Grossmann \& Lohse]{stevens2013}
{\sc \au{Stevens, R. J. A.~M.}, \au{van~der Poel, E.~P.}, \au{Grossmann, S.} \& \au{Lohse, D.}} \yr{2013}  \at{{The unifying theory of scaling in thermal convection: the updated prefactors}}.  \jt{Journal of Fluid Mechanics}  \bvol{730},  \pg{295--308}.

\bibitem[Stevenson(1979)]{stevenson1979}
{\sc \au{Stevenson, D.~J.}} \yr{1979}  \at{Turbulent thermal convection in the presence of rotation and a magnetic field: A heuristic theory}.  \jt{Geophysical \& Astrophysical Fluid Dynamics}  \bvol{12}~(1),  \pg{139--169}.

\bibitem[Straughan(2013)]{straughan2013energy}
{\sc \au{Straughan, B.}} \yr{2013} {\em {The energy method, stability, and nonlinear convection}\/}, ,  \vol{vol.~91}.  \publ{Springer Science \& Business Media}.

\bibitem[Tilgner(2022)]{tilgner2022}
{\sc \au{Tilgner, A.}} \yr{2022}  \at{Bounds for rotating rayleigh--b{\'e}nard convection at large prandtl number}.  \jt{Journal of Fluid Mechanics}  \bvol{930},  \pg{A33}.

\bibitem[Tobasco {\em et~al.\/}(2018)Tobasco, Goluskin \& Doering]{Tobasco2018}
{\sc \au{Tobasco, I.}, \au{Goluskin, D.} \& \au{Doering, C.~R.}} \yr{2018}  \at{{Optimal bounds and extremal trajectories for time averages in nonlinear dynamical systems}}.  \jt{Physics Letters A}  \bvol{382}~(6),  \pg{382--386},  \arxiv{arXiv: 1705.07096}.

\bibitem[Veronis(1959)]{veronis1959}
{\sc \au{Veronis, G.}} \yr{1959}  \at{Cellular convection with finite amplitude in a rotating fluid}.  \jt{Journal of Fluid Mechanics}  \bvol{5}~(3),  \pg{401--435}.

\bibitem[Vorobieff \& Ecke(2002)]{vorobieff2002}
{\sc \au{Vorobieff, P.} \& \au{Ecke, R.~E.}} \yr{2002}  \at{Turbulent rotating convection: an experimental study}.  \jt{Journal of Fluid Mechanics}  \bvol{458},  \pg{191--218}.

\bibitem[Wang {\em et~al.\/}(2020)Wang, Lohse \& Shishkina]{Wang2020}
{\sc \au{Wang, Q.}, \au{Lohse, D.} \& \au{Shishkina, O.}} \yr{2020}  \at{Scaling in internally heated convection: a unifying theory}.  \jt{Geophysical Research Letters}  \bvol{47},  \pg{e2020GL091198}.

\bibitem[Wang(2007)]{wang2007asymptotic}
{\sc \au{Wang, X.}} \yr{2007}  \at{{Asymptotic behavior of the global attractors to the Boussinesq system for Rayleigh-B{\'e}nard convection at large Prandtl number}}.  \jt{Communications on Pure and Applied Mathematics}  \bvol{60}~(9),  \pg{1293--1318}.

\bibitem[Wang \& Whitehead(2013)]{wang2013}
{\sc \au{Wang, X.} \& \au{Whitehead, J.~P.}} \yr{2013}  \at{A bound on the vertical transport of heat in the ‘ultimate’ state of slippery convection at large prandtl numbers}.  \jt{Journal of Fluid Mechanics}  \bvol{729},  \pg{103–122}.

\bibitem[Whitehead \& Doering(2011{\natexlab{{\em a\/}}})]{Whitehead2011}
{\sc \au{Whitehead, J.~P.} \& \au{Doering, C.~R.}} \yr{2011{\natexlab{{\em a\/}}}}  \at{{Internal heating driven convection at infinite Prandtl number}}.  \jt{Journal of Mathematical Physics}  \bvol{52}~(9),  \pg{093101}.

\bibitem[Whitehead \& Doering(2011{\natexlab{{\em b\/}}})]{Whitehead2011prl}
{\sc \au{Whitehead, J.~P.} \& \au{Doering, C.~R.}} \yr{2011{\natexlab{{\em b\/}}}}  \at{{Ultimate state of two-dimensional Rayleigh--B\'enard convection between free-slip fixed-temperature boundaries}}.  \jt{Physical Review Letters}  \bvol{106}~(24),  \pg{244501}.

\bibitem[Whitehead \& Doering(2012)]{Whitehead2012}
{\sc \au{Whitehead, J.~P.} \& \au{Doering, C.~R.}} \yr{2012}  \at{{Rigid bounds on heat transport by a fluid between slippery boundaries}}.  \jt{Journal of Fluid Mechanics}  \bvol{707},  \pg{241--259}.

\bibitem[Yan(2004)]{Yan2004}
{\sc \au{Yan, X.}} \yr{2004}  \at{{On limits to convective heat transport at infinite Prandtl number with or without rotation}}.  \jt{Journal of Mathematical Physics}  \bvol{45}~(7),  \pg{2718--2743}.

\end{thebibliography}

\end{document}